\documentclass[preprint,authoryear,11pt]{elsarticle}

\usepackage{epsfig}
\usepackage{rotating}
\usepackage{amssymb}
\usepackage{natbib}
\journal{Icarus}

\begin{document}

\begin{frontmatter}



\title{Detection of CO and HCN in Pluto's atmosphere with ALMA}


\author{E. Lellouch$^1$, M. Gurwell$^2$, B. Butler$^3$, T. Fouchet$^1$, P. Lavvas$^4$, D.F. Strobel$^5$, B. Sicardy$^1$,
A. Moullet$^6$, R. Moreno$^1$, D. Bockel\'ee-Morvan$^1$, N. Biver$^1$, L. Young$^7$, D. Lis$^{8}$, J. Stansberry$^{9}$, 
A. Stern$^7$, H. Weaver$^{10}$, E. Young$^7$, X. Zhu$^{10}$, J. Boissier$^{11}$ 
}
\address{$^1$ LESIA, Observatoire de Paris, PSL Research University, CNRS, Sorbonne Universit\'es, UPMC Univ. Paris 06, Univ. Paris Diderot, Sorbonne Paris Cit\'e, 5 place Jules Janssen, 92195 Meudon, France; emmanuel.lellouch@obspm.fr}
\address{$^2$ Harvard-Smithsonian Center for Astrophysics, Cambridge, MA 02138, USA}
\address{$^3$ National Radio Astronomy Observatory, Socorro, NM 87801, USA}
\address{$^4$ GSMA, Universit\'e Reims Champagne-Ardenne, 51687 Reims Cedex 2, France}
\address{$^5$ Department of Earth and Planetary Sciences and Physics and Astronomy, Johns Hopkins University, Baltimore, MD 21218, USA}
\address{$^6$ National Radio Astronomy Observatory, Charlottesville, VA 22902, USA}
\address{$^7$ Southwest Research Institute, Boulder, CO 80302, USA}
\address{$^{8}$ LERMA,  Observatoire de Paris, PSL Research University, CNRS, Sorbonne Universit\'es, UPMC Univ. Paris 06,75014 Paris, France}
\address{$^9$ Space Telescope Science Institute, Baltimore, MD 21218, USA}
\address{$^{10}$ Space Exploration Sector, Applied Physics Laboratory, Johns Hopkins University, Laurel, MD 20723, USA}
\address{$^{11}$ IRAM, 38400 Saint-Martin-d'H\`eres, France}


\newpage
\begin{abstract}
Observations of the Pluto-Charon system, acquired with the ALMA interferometer on June 12-13, 2015,
have led to the detection of the CO(3-2) and HCN(4-3) rotational
transitions from Pluto (including the hyperfine structure of HCN), providing a strong confirmation of the presence of CO, and the first observation of HCN in Pluto's atmosphere. The CO and HCN lines probe Pluto's atmosphere up to $\sim$450 km and $\sim$900 km
altitude, respectively, with a large contribution due to limb emission. The CO detection yields (i) 
a much improved determination of the CO mole fraction, as 515$\pm$40 ppm for a 12 $\mu$bar surface pressure 
(ii) strong constraints on Pluto's mean atmospheric dayside temperature profile over $\sim$50-400 km, with clear evidence for a well-marked temperature decrease (i.e., mesosphere) above the 30-50 km stratopause and a best-determined temperature of 70$\pm$2 K at 300 km, somewhat lower than previously estimated from stellar occultations (81$\pm$6 K), and in agreement with recent inferences from New Horizons / Alice solar occultation data. The HCN line shape implies a high abundance of this species in the upper
atmosphere, with a mole fraction $>$ 1.5$\times$10$^{-5}$ above 450 km and a value of 4$\times$10$^{-5}$ near 800 km. Assuming HCN at saturation, this would require a warm ($>$92 K) upper atmosphere layer; while this is not ruled out by the CO emission, it 
is inconsistent with the Alice-measured CH$_4$ and N$_2$ line-of-sight column densities. Taken together, the large
HCN abundance and the cold upper atmosphere imply supersaturation of HCN to a degree (7-8 orders of magnitude) hitherto unseen in
planetary atmospheres, probably due to a lack of condensation nuclei above the haze region and the slow kinetics of condensation at the low pressure and temperature conditions of Pluto's upper atmosphere.
HCN is also present in the bottom $\sim$100 km of the atmosphere, with a 10$^{-8}$ - 10$^{-7}$ mole fraction; this
implies either HCN saturation or undersaturation there, depending on the precise stratopause temperature. The HCN 
column is (1.6$\pm$0.4)$\times$10$^{14}$ cm$^{-2}$, suggesting a surface-referred vertically-integrated 
net production rate of $\sim$2$\times$10$^{7}$ cm$^{-2}$s$^{-1}$. 
Although HCN rotational line cooling affects Pluto's atmosphere heat budget, the amounts determined in this study
are insufficient to explain the well-marked mesosphere and upper atmosphere's $\sim$70 K temperature, which
if controlled by HCN cooling would require HCN mole fractions of (3--7)$\times$10$^{-4}$ over 400-800 km. We finally
report an upper limit on the HC$_3$N column density ($<$ 2$\times$10$^{13}$ cm$^{-2}$) and on the HC$^{15}$N / HC$^{14}$N
ratio ($<$ 1/125). 

\end{abstract}


\end{frontmatter}


\newpage

\section{Introduction }
Ground-based and New Horizons observations have shown Pluto's tenuous, N$_2$-dominated, atmosphere to be alluringly
Mars- and Titan- like. The long-term monitoring of stellar occultations \citep[][and references therein]{elliot03,sicardy03,olkin15, sicardy16}  has revealed a factor-of-three pressure increase in the
last 25 years, with a current pressure in the range $\sim$10-14 $\mu$bar \citep{gladstone16,sicardy16}.  This variability reflects the existence of Mars-like seasonal cycles, in which volatiles, CH$_4$ and CO in addition to N$_2$ \citep{owen93,doute99,grundy13},
are shared between atmospheric and surface ice reservoirs through sublimation/condensation exchanges and 
migration \citep{young13,hansen15,olkin15}. Most spectacularly, New Horizons discovered a thick ice cap at low latitudes (informally named Sputnik Planum) as being the main reservoir of volatiles, in addition to thinner N$_2$ frosts at mid-northen latitudes and more widespread CH$_4$ frost \citep{stern15,grundy16}. 

Near-infrared spectroscopy \citep{young97,lellouch09,lellouch15a} revealed CH$_4$ as the second most abundant atmospheric species ($\sim$0.5~\% of N$_2$), providing an explanation for the relatively warm stratosphere ($\sim$105~K-110~K at 20-50 km altitude compared to $\sim$35-55 K at the surface), according to the mechanism first described by \citet{yelle89}. Similar to Titan, the coupled photochemistry of N$_2$ and CH$_4$ is 
expected to result in the production of a suite of minor species, including hydrocarbons and nitriles \citep{lara97,summers97,
krasno99,gladstone15} and ultimately
 haze formation. Several hydrocarbons (C$_2$H$_2$, C$_2$H$_4$ and C$_2$H$_6$) were indeed detected by the New Horizons
UV experiment \citep{gladstone16}. Most impressively, extensive, optically thin hazes are seen in New Horizons images, extending to altitudes of $>$200 km, with a distinctly layered structure that may be the sign of gravity wave activity,
evidence for which is also present in stellar occultation data \citep[e.g.][]{toigo15}. 
Other key atmospheric results from New Horizons include detailed temperature profiles below 50 km, and evidence for a cold upper atmosphere ($\sim$70 K above 500 km) which effectively limits the escape rate of Pluto's atmosphere to space to much smaller values than anticipated \citep{gladstone16}.

Besides CH$_4$, the search for minor species in Pluto's atmosphere from the ground has proven difficult. A number
of searches for CO were conducted at radio-wavelengths, but as will be detailed below, results were unconstraining 
\citep{barnes93}, tentative \citep{bock01}, 
or erroneous in the case of \citet{greaves11}, 
as demonstrated by \citet{gurwell14}. In the near-IR, from IRTF/CSHELL observations of the CO(2-0) band at 2.3~$\mu$m,  
\citet{young01} determined a coarse upper limit to the CO column density ($<$ 45-130 cm-am, i.e. typically 20-60 \% of N$_2$). The same spectral region was re-observed with much improved sensitivity with VLT/CRIRES \citep{lellouch11}. Besides the undisputable presence of CH$_4$ features, the spectrum showed indication for a number of CO lines, which upon co-adding, yielded evidence for CO in Pluto's atmosphere at the 6 $\sigma$ level, 
with a 500$^{+1000}_{-250}$ ppm mixing ratio (i.e. a $\sim$0.1 cm-am column density), which within (large) error bars, was consistent with the CO:N$_2$ ratio in Pluto ice derived by \citet{doute99}.  

While most of the simple hydrocarbons found in outer solar system atmospheres are non-polar, nitriles exhibit strong rotational lines and are thus well suited
to mm/submm searches, as demonstrated notably in the case of Titan's atmosphere by the routine observation of HCN, HC$_3$N, CH$_3$CN and some of their isotopes, and the recent detection of ethyl cyanide (C$_2$H$_5$CN) and vinyl cyanide (C$_2$H$_3$CN) with ALMA
\citep{cordiner15a,cordiner15b}.  
In this paper, we report on searches for CO, HCN and HC$_3$N in Pluto's atmosphere with ALMA, leading
to the strong confirmation of CO and the first detection of HCN.

\section{Observations and data reduction}
Observations of the Pluto-Charon system were obtained with the 12-m 
array of the Atacama Large Millimeter Array (ALMA).  This synthesis
array is a collection of radio antennas, each 12 m in diameter, deployed
out on the Chajnantor Plateau in the high northern Chilean Andes.  Each 
pair of antennas acts as a two-element interferometer, and the
combination of all of these individual interferometers allows for the
reconstruction of the full sky brightness distribution 
\citep{thompson01}.

ALMA is currently tunable in 7 discrete frequency bands, from $\sim$90 to $\sim$950 GHz.  
Observations reported here were taken in Band 7, near 350
GHz.  Scientific goals  included: (i) independent 
measurements of continuum emission from the surfaces of Pluto and Charon and
(ii) search for emission from atmospheric constituents (CO, HCN and HC$_3$N).  
To achieve these purposes, we tuned the receivers and set up the correlator in four separate frequency ranges
(spectral windows) and resolutions: one ``continuum'' (so-called ``Time Division Mode" -- TDM) range,
and three ``spectral line'' (so-called ``Frequency Division Mode" -- FDM) ranges. Details on the frequency ranges,
spectral resolutions and targetted lines are given in Table~\ref{spectral_setup}. 
 
\begin{table}
\caption{ \label{spectral_setup}
   Observation spectral set-ups.
   }
\vspace{5truemm}
\centering
\footnotesize
\begin {tabular}{cccc}
\hline\hline
\noalign{\vspace{3pt}}
\# & Spectral range (GHz)  & Resolution (MHz) & Targetted lines and frequencies (GHz)\\
\noalign{\vspace{3pt}}
\hline
\noalign{\vspace{2pt}}
0 & 354-356 & 15.625 & Continuum  \\
\hline
1 & 345.598--345.832 & 0.141 & CO(3-2); 345.796   \\
  &                  &       & HC$_3$N(38-37); 345.609   \\
\hline
2 & 354.380--354.849 & 0.282 & HCN(4-3); 354.505   \\
  &                  &       & HC$_3$N(39-38); 354.697   \\
\hline
3 & 344.011--344.949 & 0.488 & HC$^{15}$N(4-3); 344.200    \\
\noalign{\vspace{2pt}}
\hline\hline
\end{tabular}
\end{table}

All observations were taken in dual-linear polarization.
As atmospheric line emission is not expected
to be polarized, data in the two polarizations were combined to provide a measurement of the
intensity (Stokes I).

The observations were undertaken on two separate dates, June 12
and June 13, 2015. There were 39 operating antennas in the array, in the
C36-7 configuration.  This configuration, which has a maximum antenna
separation of $\sim$ 800 m, yielded at these frequencies a
$\sim$0.35$^{\prime\prime}$ resolution (i.e., 8100 km at Pluto's distance).  
This is sufficient to
separate the emission of Pluto from Charon, but not enough to 
provide spatial resolution on the surface of these bodies.
Table \ref{obs-tab} provides relevant
geometrical parameters for the two observing dates. 

On each date, the entire observation took 1 hour 40 minutes, of
which about 55 minutes were on Pluto/Charon.  We observed the calibrator
J1924-2914  to calibrate bandpass and delay, and Titan for absolute flux
density calibration.  The expected overall flux density scale accuracy
is 5\%, using a standard model for Titan \citep{butler12}.  We used
the point-like radio source J1911-2006  to calibrate the phase of the
atmosphere and the antennas as a function of time, as well as temporal
variations of amplitude gain; the derived flux density
of this calibrator was $\sim$0.75 Jy on both observing dates.

Initial calibration of the data was provided by the ALMA observatory,
and done in the CASA reduction package, version 4.3.1
\citep{muders14}.  The actual measured quantity of an
interferometer like ALMA is a sampling of the complex visibility
function for the baselines (length and orientation) between each of its antennas.
The visibility function is the two-dimensional Fourier transform of the
sky brightness distribution.  The individual samples of the visibility
function are referred to as visibilities, and are complex quantities
(real and imaginary, or amplitude and phase).  After the initial
calibration, the data product was a set of visibilities for each of the
observing dates.

Data acquisition with ALMA is performed with fixed frequencies.  Because of
this, a correction to the data must be made to account for the changing
Doppler shift of Pluto/Charon with respect to ALMA.  In addition, this
correction puts the frequency scale in the proper rest frame.  This
correction was done with the CASA tasks {\it fixplanets()} and {\it
cvel()}.  These tasks were run on each of the above spectral windows
separately.

At this point we exported the data from CASA and continued the data
reduction in the AIPS package
(\verb?www.aips.nrao.edu/CookHTML/CookBook.html?).  We imaged the
continuum spectral window, from which we used the technique of self-calibration
to correct for residual phase errors (left over after
the removal of the time-variable complex gain from the calibrator) in
the visibilities \citep{cornwell99}.  We did this iteratively with
solution intervals of 5, 2, and 1 minute.  Since the phase errors at
this point are primarily due to tropospheric water vapor, and the frequency difference
between the spectral windows is not large, the phase solutions found for
the continuum window could be used to correct the data in the
spectral line spectral windows.

We then subtracted the continuum in the three spectral line windows (18.2 mJy in CO(3-2) and HC$^{15}$N(4-3),  
and 18.9 mJy in HCN(4-3)).
We did this in the visibility domain, using an average of spectral
channels that were far outside the line emission region
\citep{vanlange90}. Finally, averaging continuum-subtracted visibilities for each spectral channel
yielded the spectra in each of the target lines. Strictly speaking, these spectra correspond
to the flux received at the central spatial position (i.e., pointing to Pluto) and not to the total
emitted flux. Given the relative size of the source ($\sim$0.15" for Pluto's diameter at 600 km altitude)
and the beam ($\sim$0.35" HPBW), convolution effects amount to a typical 10 \% flux loss.  

While no atmospheric emission was detected from Charon, the main highlight
of the observations was the detection of line emission due to CO and HCN
from Pluto's atmosphere. Following preliminary reports of these data
\citep{lellouch15b,lellouch15c,gurwell15}, this result is the focus
of the present paper. Continuum measurements 
have been briefly presented \citep{butler15} and will be described in detail
in a future publication.

\vspace{1truecm}

\begin{table}
\caption{ \label{obs-tab}
   Observing dates and geometry.
   }
\vspace{5truemm}
\centering
\footnotesize
\begin {tabular}{ccccc}
\hline\hline
\noalign{\vspace{3pt}}
Date & Time range (UTC) & Geocentric & Sub-Earth &
Pluto-Charon \\
      &                  &       Distance (AU)         & Longitude$^a$ &
separation (arcsec)\\
\noalign{\vspace{3pt}}
\hline
\noalign{\vspace{2pt}}
2015-Jun-12 & 04:15--05:55 & 31.958 & 157 & 0.70 \\
2015-Jun-13 & 03:25--05:05 & 31.952 & 103 & 0.84 \\
\noalign{\vspace{2pt}}
\hline\hline
\multicolumn{5}{l}{\footnotesize $^a$ Sub-observer East longitude at mid-point. We adopt the new IAU convention} \\
\multicolumn{5}{l}{\footnotesize for definition of the North Pole \citep{buie97,zangari15}, with current}\\ 
\multicolumn{5}{l}{\footnotesize summer in the northern hemisphere. Zero longitude on Pluto is the sub-Charon}\\ 
\multicolumn{5}{l}{\footnotesize point and the sub-observer point longitude decreases with time.}

\end{tabular}
\end{table}

\section{Data assessment}
Continuum-subtracted Pluto spectra in the vicinity of the CO(3-2) and HCN(4-3) lines are shown in Fig. 1 separately for the
two observing dates. A strong detection of both lines was achieved on each day, providing the first detection
of HCN in Pluto's atmosphere and the first observation of CO at radio-wavelengths. The HCN spectra clearly
show the hyperfine structure, with weak emission features at -1.6 and +2.0 MHz from the main line. These
secondary emissions (termed hereafter the ``satellite lines") arise from transitions with very similar
lower energy levels but with absolute strengths (hence opacities) $\sim$45 times weaker than the main line (which is
itself the unresolved sum of three components). As shown below,
these secondary lines are quite useful for constraining the HCN vertical distribution.

Data from the two different days do not show significant differences. In particular, although observations
on June 12 (subearth longitude L = 157$^{\circ}$E) fully encompassed Sputnik Planum -- centered at 20$^{\circ}$N, 175$^{\circ}$E and which is the region showing the largest concentration of CO ice \citep{grundy16}\footnote{ \citet{schmitt16} report that CO ice is also present at mid-northern latitudes in Venera Terra and Burney crater.} -- this does not result in 
a significant enhancement of the gas CO feature; gaussian fits of the CO line indicate peak contrasts of
58.1 and 62.8 mJy for June 12 and 13, respectively, i.e. a 1-$\sigma$ difference given that the rms noise
of the individual spectra is 4.6 mJy. This is qualitatively consistent with expectations from recent multi-species 
General Climate Models that CO is well mixed with N$_2$ \citep{forget16}. Data from the two days were therefore averaged for further analysis.

The averaged CO(3-2) and HCN(4-3) line contrasts are 61 mJy and 99 mJy at $\Delta{\nu}$ = 141 and 282 kHz resolution, respectively.
With 1-$\sigma$ rms residuals of 3.28 mJy and 2.85 mJy, this gives a S/N detection of 19 for CO and 35 for HCN at the peak.
However, the CO and HCN line shapes are very different. The main HCN line is narrow ($\sim$0.8 MHz FWHM), with residual
flux falling practically to zero between the main and the satellite lines. The CO line is broader ($\sim$2 MHz FWHM),
with signal being detected up to $\pm$5 MHz in the wings. The S/N on the integrated line is 160 for CO and 58 for HCN.

Finally, no detection of HC$_3$N or HC$^{15}$N was achieved, with 1-$\sigma$ rms residuals of 3.28 mJy at $\Delta{\nu}$ = 141 kHz
for HC$_3$N(38-37), 2.85 mJy at $\Delta{\nu}$ = 282 kHz for HC$_3$N(39-38) and 1.88 mJy for HC$^{15}$N(4-3) at R = 488 kHz.

While data-model comparisons are hereafter presented in the observational flux units, expressing the data
in brightness temperature is useful for qualitative assessment and comparison with previous CO searches. With a Pluto apparent
size of 0.103 arcsec, the CO line contrast of 61 mJy, superimposed to a 18.2 mJy continuum, translates into a 119 K
brightness temperature (T$_B$) at the peak, and a 85 K Rayleigh-Jeans temperature contrast ($\Delta T_{RJ}$) with
respect to the continuum. For HCN, the 99 mJy contrast, superimposed to 18.9 mJy continuum, indicates a peak T$_B$ = 165 K and $\Delta T_{RJ}$~=~132 K. The fact
that the peak T$_B$ are higher than Pluto's stratopause temperature directly implies that a significant fraction
of the emission originates from limb viewing (i.e. rays not intercepting the surface), enhancing the size of Pluto's emitting disk. 

Based on IRAM-30 m observations in April 2000, \citet{bock01} reported a tentative detection of the CO(2-1) line, with an integrated line area of (18$\pm$4) mK-MHz and a $\sim$15 mK main-beam temperature contrast. With a 10.6 arcsec half-power beam width (HPBW) of the telescope at CO(2-1) and a Pluto 0.111'' apparent size at the time, this line contrast would have corresponded 
to a $\Delta T_{RJ}$ = 196 K contrast at Pluto. This is inconsistent with $\Delta T_{RJ}$ = 85 K \footnote{Strictly speaking, this comparison should include a $\sim$10 \% upwards correction of
this value to correct for beam effects, given that the IRAM-30 m observations are truly unresolved.} that we measure
for CO(3-2) (all the more so when considering that CO(3-2) is also intrinsically stronger than CO(2-1)), implying that the possible
line reported by the authors, but wisely treated as an upper limit, was not real. Similarly but more detrimental
for what was published as a ``discovery", the claimed, yet unmodelled, detection of CO(2-1) by \citet{greaves11}
from JCMT, with a line contrast $>$4 larger (i.e., $\sim$800 K contrast) than that in \citet{bock01}, was undoubtedly spurious. 
Note that based on a deep search (17 hours integration) for CO(2-1) with the SMA, \citet{gurwell14} reported no
detection while reaching a 82 mJy rms 
at 100 kHz resolution; as such this was already enough to refute the \citet{greaves11} results at 10-$\sigma$ confidence.

\section{Modelling and analysis}

To analyze the data, we developed our own radiative transfer code, fully accounting for the spherical geometry associated with the large extent of Pluto's atmosphere compared to the object's radius. The model covers the 0-1300 km altitude
range by means of 181 layers, with an altitude-varying layer thickness. Line opacities for CO and HCN were calculated using line parameters from the Cologne database \citep{muller01}, including the hyperfine structure for HCN. Collisional broadening by N$_2$ were taken from \citet{koshelev09}, \citet{priem00} and \citet{yang08}, giving HWHM's of 2.14 and 4.20 GHz/bar at 300 K for CO(3-2) and HCN (4-3), with temperature exponents of --0.84 and --0.69 respectively. Emission from the surface was calculated by using a Fresnel-emission model with a dielectric constant of 3.5, adjusting the surface temperature to match the measured continuum flux (T$_{surf}$~=~38.4 K and 38.0 K for CO and HCN respectively)\footnote{Other combinations of dielectric constant and surface temperature are possible, with no impact on the atmospheric results.}. Atmospheric pressure-temperature-altitude profiles were calculated using a molecular mass of 27.94 to account for the presence of CH$_4$ in Pluto's atmosphere. The altitude variation of the gravity was accounted for, using a surface radius of 1190 km based New Horizons/REX \citep{gladstone16}.  A surface pressure of 12 $\mu$bar was nominally assumed, intermediate between values inferred from the June 29, 2015 stellar occultation \citep[11.9--13.7 $\mu$bar,][]{sicardy16} and from the New Horizons/REX \citep[11$\pm$1 and 10$\pm$1 $\mu$bar at entry and exit, respectively,][]{gladstone16}. As discussed below we also considered values of 10 and 14 $\mu$bar. Contributions to the emission of the individual (vertical and horizontal) lines-of-sight were calculated as a function of distance D from Pluto's center (using geometrically exact pathlengths in each layer) for 0~km~$<$~D~$<$~2500~km, convolved with the synthetic 0.35" HPBW beam of the array 
and finally co-added. 

Local thermodynamical equilibrium (LTE) was assumed throughout the atmosphere when calculating the HCN / CO rotational populations and performing the radiative
transfer calculations. The de-excitation coefficient for the HCN(4-3) transition by collisions with N$_2$ can be estimated to
2.2$\times$10$^{-11}$ cm$^3$s$^{-1}$ at 30-100 K. This value comes from (i) an excitation coefficient of 9$\times$10$^{-12}$ 
cm$^3$s$^{-1}$ for HCN(4-3) by He \citep{dumouchel10}\footnote{See http://basecol.obspm.fr.} (ii) its scaling by
a factor 3.1 to obtain the excitation rate by N$_2$, following \citet{rezac13} (iii) the application of
detailed balance relating excitation and de-excitation coefficients. With an Einstein coefficient A = 2.05$\times$10$^{-3}$  
s$^{-1}$, rotational LTE is ensured for HCN(4-3) up to a critical density of 9.3$\times$10$^{7}$ cm$^{-3}$, i.e.
a pressure of 9$\times$10$^{-13}$ bar. In Pluto's atmosphere this is reached at an altitude of $\sim$1150 km, where as we show
below, the contribution to the total HCN emission is very small. For CO(3-2), due primarily to an Einstein coefficient
820 times weaker (and a de-excitation coefficient $\sim$2 times stronger), the LTE domain extends even much higher.

\subsection{CO and temperature retrievals}
\subsubsection{Method}

The CO line contrast and shape depend on both the CO mole fraction and Pluto's mean temperature profile.
In all our retrievals, we assumed that CO is vertically  uniformly mixed. This assumption is supported by (i) the absence of gravitational separation between N$_2$ and CO and (ii) the stability of CO against photolysis. Note also that general circulation models indicate that CO is horizontally and vertically well mixed with N$_2$ \citep{forget16}. For temperature retrieval we used a constrained and regularized algorithm following methods detailed in 
\citet{conrath98} and \citet{rodgers00}, see also \citet{fouchet16} for a recent application. In this approach, temperature profiles are initialized to an a priori profile, and are constrained to stay close to this a priori at levels where the measurements contain no information. In pressure regions where information is available, the departure from the a priori profile is regularized (i.e. smoothed to some vertical resolution) to avoid spurious oscillations of the output profiles. In practice we set the correlation matrix $\bf{S}$ to be a gaussian function with a correlation length equal to the atmospheric scale height. As the inversion process involves a linearization of the radiance with respect to temperature\footnote{For practical simplicity, the Jacobian matrix $\bf{K}$ was calculated by taking into account only the derivative of the Planck function with temperature}, it is an iterative process. Starting from an initial temperature profile {\bf T} ($n$ levels), the inversion process returns a vector {\bf $\Delta$T} which must be added to {\bf T} for next iteration, and is given by:

\begin{displaymath}
\bf{\Delta T} = \bf{U} \Delta I 
\end{displaymath}
\begin{displaymath}
\textrm{where}~~~~\bf{U} =  \alpha S K^T (\alpha K S K^T + E)^{-1} 
\end{displaymath}

Here {\bf $\Delta$I} is the difference between the observed and modelled spectral radiances ($m$ values). $\bf{K}$ is the Jacobian matrix, i.e. the $m$ x $n$ matrix of derivatives of the radiances with temperatures K$_{ij}$= dI$_i$/dT$_j$ where I$_i$ is the radiance at frequency $i$ and T$_j$ is the temperature at level $j$. $\alpha$ is a weighting factor and $\bf{E}$, the error covariance matrix of the measurements, is simply taken as a diagonal matrix with all non-zero elements equal to the square of the uncertainty of the measured radiance. In essence, $\alpha$ represents the relative weights to give to
the measurements and to the a priori at each step of the iteration process. Following \cite{fouchet16}, we found that
an optimum choice of $\alpha$ (in terms of stability and rapidity of the algorithm) was achieved by equalling the traces
of the $\alpha${\bf KSK$^T$} and {\bf E} matrices.  Iterations were performed until convergence -- defined by the constancy of the fit quality (rms between observations and models) within 1 \% from an iteration to the next one -- was reached. 
 
As explained e.g. in \citet{rodgers00} and \citet{fouchet16}, the information content (in terms of sounded altitude and vertical resolution)  is characterized by the averaging kernels $n$ x $n$ matrix $\bf{A}$, defined by $\bf{A}$ = $\bf{UK}$, which contains
the partial derivatives of the retrieved state with respected to the true state vector (here, the temperature profile). In particular, the sum of
all values in a given row $k$ of matrix $\bf{A}$ is an indicator of the relative weight of the measurements and of the a priori in determining the temperature at level $k$, and the number of independent temperatures that can be overall retrieved from the data is given by  Tr($\bf{A}$) = $\sum\limits_{i=1}^n$ A$_{ii}$ \citep[][see pp. 30-31 and 46-47]{rodgers00}. 

In a first step, we inverted the CO line with a fixed CO mole fraction of 500 ppm, the central value suggested by \citet{lellouch11}. Since the observed CO line, and particularly line wings, is also sensitive to the CO mole fraction, we then also performed simultaneous temperature/abundance retrievals. For this, we followed the formalism of \citet[][his Equations 20-24]{conrath98}, though with the simplification associated with the mole fraction vertical uniformity. To calculate the derivative of the spectral radiance I$_i$ with respect to the CO abundance, we simply calculated two spectra with CO mole fractions (q) differing by 10 \%, from which 
dI$_i$/d(ln~q) was inferred. As detailed hereinafter, these ``fixed-CO" and ``free-CO" inversion processes were run with a variety
of a priori temperature profiles. In all cases, the best fit corresponded to a reduced $\chi^2$ of $\sim$0.95 over a $\pm$15 MHz
interval centered on the CO(3-2) line frequency, demonstrating the ability of the model to reproduce the observations.

\subsubsection{Results, information content and uncertainties}
We started with the nominal temperature profile derived from stellar occultations by \citet{dias15}. 
Fig. 2 (green dashed line) clearly shows that the a priori temperature profile which exhibits a $\sim$81 K temperature above 200 km, leads to an large overestimation of the line core contrast. This effect can be corrected by modifying the
upper atmosphere (above 200 km) temperatures to reach an asymptotic value of 69~K, leaving the atmosphere below
200 km untouched (``cold Dias-Oliveira", or ``DO15 + 69 K" profile). The resulting profile (blue line) matches the line contrast, but the shape of the line core is not completely fit, as the model predicts weak absorptions at $\pm$0.4 MHz not observed in
the data, which are produced in the model by too strong a thermal gradient over 100-300 km. 
Fixed-CO and free-CO inversions were then performed, using the original stellar occultation profile as a priori.
For free-CO inversions, we specified an a priori CO mole fraction of 500 ppm, checking that results (i.e. the output thermal profile and CO mole fraction) were independent on this assumption. 
In both cases, the inversion process returns a temperature of $\sim$65 K at 350 km, progressively relaxing towards the a priori at higher altitudes. In addition, the returned temperature profile is colder than the a priori over 50-200 km, by up to 5-10~K.
The fit improvement over the ``cold Dias-Oliveira" profile is at the 4~$\sigma$ level.
When the CO mole fraction was left as a free parameter, the inversion process returned a 507$\pm$21 ppm value.  The associated CO column density is (0.099$\pm$0.004) cm-am. In vertical viewing (disk center) and at line center, the total optical depth $\tau$ is equal to 119 and the $\tau$ = 1 level is reached at 300 km. In the linewings, unit optical depth at the surface is reached at $\pm$1.2 MHz from line center.  

We extended these retrievals by using several alternative a priori profiles, namely: (i) the ``standard" profile derived from New Horizons  \citep[see Fig. 3 from][]{gladstone16}, which combines results from Alice (above 200 km) and REX (below 60 km) with expectations based on radiative-convective thermal and diffusion models for N$_2$/CH$_4$ (ii) isothermal profiles at 50 K, 70 K and 100 K. Although these uniform profiles are obviously not realistic given pre-existing knowledge of Pluto's atmosphere, they provide useful ``end-member" cases to assess the robustness of the results. Results are given in Table 3 and Fig. 3.

Despite the large diversity of a priori profiles, all retrieved thermal profiles consistently show a common feature, namely a steady temperature decrease over at least 100-350 km. For the two most plausible a priori profiles, the stellar occultation and the New Horizons observations, the range of altitude with negative temperature gradient extends over $\sim$30-400 km. Temperatures over 250-400 km are particularly stable against a change of a priori, with a best determined temperature of 70$\pm$2 K at 300 km, in full agreement with the
New Horizons / Alice evidence for a cold upper atmosphere \citep{gladstone16}. The other strongly robust result is the determined CO mole fraction in the ``CO-free retrievals", which for the five a priori considered, varies from 474$\pm$25 ppm to 523$\pm$20 ppm. Restricting ourselves
to the two most realistic a priori profiles, the choice of the a priori induces a $\pm$1.5 \% uncertainty about a central 515 ppm
value.

These results are in line with expectations on the information content based on examination of the averaging kernels. 
Averaging kernels at 0, 100, 200, 300, 500 and 700 km are shown in Fig. 4. The typical width of these kernels, 200 km,
is an estimate of the vertical resolution. The peak altitude of these kernels extends up $\sim$350 km, indicating that the
information content decreases above this level. The total kernel (i.e. the sum of all values in a given row of the averaging
kernel matrix) as a function of altitude exceeds 0.8 over 170-380 km, indicating the clear dominance of the measurements
themselves over the a priori in constraining the temperatures there. The number of independent temperature measurements,
as estimated from the trace of the averaging kernel matrix, is $\sim$2.5.

Errors on the retrieved temperature profile and CO abundance resulting from random noise in the measurements are quantified in the inversion method \citep[see Equations 23 and 24 in][]{conrath98}. These random temperature errors are of order 1-1.5 K over 100-400 km, progressively decreasing to smaller values (e.g. 0.5 K at the ground and at 500 km) outside of the region best constrained by the measurements. Naturally, outside of these regions, the true uncertainties are dominated by the 
a priori errors. Additional systematic errors originate from (i) the uncertainty in surface pressure (ii)
the flux scale accuracy of the measurements (5 \%). To investigate those, and starting from a ``mixed" a priori profile (defined
as some weighted mean of the \citet{dias15} and \citet{gladstone16} profile, where the relative weight of
the latter increases with altitude), we re-ran the inversion process
with surface pressures of 10, 12 and 14 $\mu$bar and the nominal data calibration. Additionally, for the 12 $\mu$bar case,
we considered the effect of multiplicative factors of 0.95 and 1.05 on the data. Results are given in Table 3 for the CO
abundance and in Fig. 5 for the temperature. The $\pm$5 \% calibration uncertainty results in a typical $\pm$2 K temperature uncertainty, while the effect of surface pressure on the retrieved temperatures is smaller. Conversely (Table 3), the $\pm$5~\% calibration uncertainty induces
a $\pm$6 \% error bar on the CO mole fraction, but the effect of surface pressure is much larger: we find that the retrieved
CO mole fraction essentially scales as 1/p$_{surf}^{1.9}$. The large surface pressure dependence is 
due to the fact that the CO mole fraction is mostly 
determined by the radiance in the CO linewings (i.e. at distances much larger than the Lorentz and Doppler linewidths), 
where the opacity scales as the product of the CO mole fraction by the total number density (proportional to pressure) and the
Lorentz HWHM (also proportional to pressure).
In summary, for a fixed
surface pressure, error bars on the CO mole fraction and the CO column density due to the choice of the a priori, measurement errors and calibration uncertainty amount to 
1.5 \%, 4 \%, and 6 \%, respectively, giving a global (quadratically-summed) uncertainty of 7.5~\%. We conclude that the CO mole fraction in Pluto's atmosphere is 515 $\pm$ 40 ppm for p$_{surf}$ = 12 $\mu$bar, and must be scaled by (12 $\mu$bar/p$_{surf}$)$^{1.9}$ for other adopted values of p$_{surf}$. The surface-referred column density\footnote{Throughout the paper, to account for Pluto's sphericity, we use column densities referred to the surface,  i.e. calculated as $\int$ N(z)(1+z/R$_{pl}$)$^2$dz, where z 
is altitude, R$_{pl}$ is Pluto radius and N(z) is the gas concentration at altitude z.}
is (2.69$\pm$0.21)$\times$10$^{18}$ mol cm$^{-2}$ (0.100$\pm$0.008 cm-am) for p$_{surf}$ = 12 $\mu$bar, and scales as (12 $\mu$bar/p$_{surf}$)$^{0.9}$.

\begin{table}
{\bf Table 3: Retrieved CO abundance vs model assumptions }

\vspace*{5mm}

\begin{tabular}{|c|c|c|c|c|}
\hline
p$_{surf}$  &  A priori  & Calibration & CO mole   & CO column\\
($\mu$bar) & T(z) profile & factor & fraction (ppm) &  density (cm-am) \\
\hline
12 & Isothermal 50 K & 1.00 & 474$\pm$25 &  0.091$\pm$0.005 \\
12 & Isothermal 70 K & 1.00  & 486$\pm$25 & 0.093$\pm$0.005 \\
12 & Isothermal 100 K & 1.00 & 514$\pm$26 & 0.098$\pm$0.005 \\
12 & \citet{gladstone16} & 1.00 & 523$\pm$20 & 0.101$\pm$0.004\\
12 & \citet{dias15} & 1.00 & 507$\pm$21 & 0.099$\pm$0.004 \\
12 & Mixed & 1.00 & 510$\pm$20 & 0.099$\pm$0.004 \\
14 & Mixed & 1.00 & 388$\pm$15 & 0.088$\pm$0.004 \\
10 & Mixed & 1.00 & 720$\pm$31 & 0.117$\pm$0.005 \\
12 & Mixed & 1.05 & 542$\pm$21 & 0.105$\pm$0.005\\
12 & Mixed & 0.95 & 482$\pm$19 & 0.094$\pm$0.004 \\
\hline
\end{tabular}
\end{table}

\subsection{HCN results}
\subsubsection{Choice of temperature profile}
To analyze the HCN data, we must first select a temperature profile among those fitting the CO line. As
is clear from Figs. 3 and 4, there is significant latitude in this respect in the lower atmosphere ($<$100 km),
where the retrieved thermal profile from CO is strongly influenced by the a priori. Unfortunately,
the New Horizons standard profile \citep{gladstone16} and the stellar occultation profile \citep{dias15} 
show noticeable differences there. The former, which follows closely the 
REX profiles in the first 50 km, has a much less steep gradient than the latter over 0-20 km, and as we will
show later (Section 5.1.2), would produce stellar occultation lightcurves markedly different from the observed.
In this situation, we adopted as nominal profile the one retrieved from CO using the above ``mixed" thermal profile 
as a priori (red line in Fig. 5). We also briefly studied the effect of using instead the ``cold Dias-Oliveira" profile,
which, although not optimized, led to an only slightly worse fit of the CO line. Given that the two profiles have 
different maximum temperatures (106~K and 110~K, respectively), this permitted us to study the impact of the stratopause temperature on the HCN results.

\subsubsection{HCN initial fits}
Simple hand-fitting of the HCN line using the nominal temperature profile revealed a few basic characteristics of the HCN distribution (Fig. 6). First, the narrowness of the HCN features (with no visible Lorentzian wings) and the near-zero emission level between the HCN main line and the two satellite lines exclude a uniform distribution of HCN gas in Pluto's atmosphere. For example, a constant HCN = 10$^{-7}$ mole fraction (red dashed-dotted line), while strongly under-estimating the main line contrast, clearly overpredicts the satellite lines and produces undesired linewings. Second, we considered an HCN profile following local saturation for solid-vapor equilibrium, using for that purpose the HCN vapor pressure vs temperature dependence provided by \citet{fray09}. Such a profile (green, long-dashed) is also strongly inadequate, especially underpredicting both the main and the satellite lines. Because the HCN vapor pressure is such a sensitive function of temperature, and our temperature retrievals have a global precision/accuracy of a few degrees, we next investigated the effect of enhancing the temperatures by +2 K when calculating the HCN vapor pressure. This case (shown in Fig. 6 as the pink dotted line) does match the contrast of the satellite lines, but continues to underpredict the main line (and overestimates the residual level in-between). Pursuing the exploration of HCN profiles with simple altitude dependence, we considered the case of uniform HCN distribution (with a q$_0$ mole fraction above some limiting altitude z$_0$, and zero HCN at deeper levels. A rather satisfactory fit of the entire HCN spectrum (matching, in particular, the relative contrast of the main and satellite lines) was obtained for z$_0$ = 460 km and q$_0$ = 4$\times$10$^{-5}$, corresponding to a column density referred to Pluto's surface of 1.3$\times$10$^{14}$ cm$^{-2}$. Note however that this model solution produces a zero flux minimum in-between the main and satellite HCN lines, while the data marginally indicate non-zero residual flux there. This residual flux, if real, does suggest that some of the emission originates from the atmosphere below 100 km. At this stage, we obtained an optimum manual fit of the HCN spectrum (solid light blue line in Fig. 6) with a two-component HCN distribution combining (i) a lower-atmosphere component, calculated as saturated HCN for the adopted temperature profile (with a 106 K stratopause) (ii) a uniform HCN layer with q$_0$ = 1.5$\times$10$^{-5}$ above z$_0$ = 450 km altitude. This two-component HCN profile has a surface-referred column density of 2.0$\times$10$^{14}$ cm$^{-2}$ (0.47$\times$10$^{14}$ and 1.53$\times$10$^{14}$ cm$^{-2}$ for the upper
and lower atmosphere components, respectively). In this favored model, the opacity at the main HCN line center and in vertical viewing is 3.8 (with relative contributions of 1.2 and 2.6 for the upper and lower layer respectively), and the total vertical opacity at the center of the satellite lines is equal to 0.10. 

Results in terms of HCN content and vertical distribution are very similar if the ``cold Dias-Oliveira" temperature profile 
is used instead. In Fig. 6, we additionally show the best fit ``two-component" distribution in this case. The total HCN column density is 
1.9$\times$10$^{14}$ cm$^{-2}$. The upper atmospheric component is identical to the one found for the nominal thermal profile. The only important difference is in the interpretation of the lower atmospheric component. For this thermal profile, this component
corresponds to HCN saturation at temperatures 5 K colder than prescribed by the profile, equivalent to a factor of 8 undersaturation. Thus, depending on the precise temperature in the stratopause region, we
infer that HCN is either at saturation (for a stratopause at 106 K), or undersaturated by up to one order of magnitude
if the stratopause is warmer.


\subsubsection{HCN retrievals}
We also performed some inversion retrievals of the HCN profile, returning to the nominal thermal profile (Fig. 7). Given that the hyperfine lines are optically thin and that even the main line is only moderately thick, one can expect non-unique solutions to appear in the inversion process. Such efforts are nonetheless valuable to assess the information content of the HCN observations.
Starting with a vertically constant a priori profile with HCN  = 1$\times$10$^{-5}$, the retrieved profile (Fig. 7, red solid line)
confirms the need for sharp decrease of HCN with decreasing altitude, from $\sim$5$\times$10$^{-5}$ at 800 km down to
less than 10$^{-7}$ below 150 km. In this case, the HCN mole fraction above 150 km altitude is enhanced by $>$ 6-7 orders
orders of magnitude above the saturation value. Similar HCN profiles are returned for various a priori
values of the HCN mole fraction (retrievals are also shown in Fig. 7 for initial HCN = 1$\times$10$^{-6}$ and 3$\times$10$^{-5}$).
Note that all of these profiles have significant HCN content over 200-400 km (HCN = 1$\times$10$^{-8}$ - 1$\times$10$^{-6}$)
while the manual fits in the previous subsection indicated that solutions can be also found for negligible
HCN amounts in this region. We therefore tested yet another a priori, in which the HCN input profile is the sum
of a constant 1$\times$10$^{-5}$ mole fraction above z$_0$ = 470 km and, below this altitude, of a saturated profile at a 
temperature 2 K higher than that of the nominal profile. While the a priori profile has a HCN mole fraction of
$\sim$3$\times$10$^{-7}$ at the stratopause, the inversion process returns a $\sim$2$\times$10$^{-8}$ mole fraction 
there (green curves in Fig. 7). This confirms that the contrast of the HCN satellite lines and the residual levels between them and the main line
provide an upper limit to the HCN ``stratospheric" mole fraction.

Averaging kernels for HCN are shown in Fig. 8, for the profile returned with a constant 1$\times$10$^{-5}$ a priori mole fraction.
They clearly show the two altitude regimes probed by the HCN transition (i) the 600-800 km range, probed in the 
main line and (ii) the lower atmosphere at 50-150 km, probed in the satellite lines.

The number of independent measurements, as given by the trace of the averaging kernel matrix, is $\sim$1.5. 
In fact, the data constrain rather well the HCN mole fraction to be $\sim$4$\times$10$^{-5}$, over 
600-800 km, and as explained above, puts a limit on the amount of HCN in the near-surface ($<$100 km)
atmosphere. Somehow ironically, these are the ranges where the temperature is the least well determined 
from the CO line.
The HCN column density indicated by these retrievals is in the range (1.2--1.4)$\times$10$^{14}$
cm$^{-2}$.

\subsubsection{Evidence for HCN supersaturation and search for alternate solutions}
The above manual fits and retrievals both indicate that for the nominal and alternative thermal profiles based on the CO
line, the HCN line data imply a huge ($>$7 orders of magnitude at 800 km) supersaturation of HCN, at least in Pluto's upper atmosphere ($>$ 500 km), over expectations based on the \citet{fray09} expression for the HCN vapor pressure. As vapor pressure
measurements against which this expression can be verified are not available at temperatures below 230 K, its applicability
down to $\sim$ 70 K (i.e. over $>$ 10 orders of magnitude in pressure) might seem highly hazardous. However,
in addition to the vapor pressure, the heat capacity of solid HCN has been measured from 15 K to the triple point
at $\sim$260~K, from which that of the gas can be calculated. This approach, on which the \citet{fray09}
saturation law is based, has been validated by the authors on several other atoms and molecules (e.g Ar, CO$_2$) over wide
pressure-temperature ranges (up to 15 orders of magnitude in pressure). 
This leads the authors (B. Schmitt and N. Fray, priv. comm.) to believe that uncertainties on the HCN
pressure curve should not be larger than a factor of a few (typically 2-5) over T = 65~-~100~K.
Therefore, they cannot be the cause of the apparently highly supersaturated HCN abundances, and further investigations
are warranted. 

An obvious way to alleviate the requirement for HCN supersaturation is to invoke a warm region in the upper atmosphere
in which HCN would be concentrated. Although the latter is not naturally indicated by the CO retrievals,
such a warm region might not be inconsistent with the CO line given that the latter is weakly sensitive to
temperatures above 450 km. Starting from our nominal thermal profile, we explored a suite of temperature
changes above 150 km, requiring that HCN is at saturation throughout the atmosphere. This specific 
assumption is the one that minimizes the required temperature increment in the warm layer (hence causing the least impact
to the CO line). A complication in this exercise is that the lower atmosphere ($<$ 100 km) significantly contributes to the HCN emission. For example, Fig. 6 (green line) shows that for our nominal thermal profile, a saturated HCN profile can account for approximately half of the HCN emission, both in the main and satellites lines. With that in mind, and keeping the temperature and HCN profiles fixed below 150 km, we explored
thermal profiles characterized by (i) the altitude (z$_1$) and (ii) temperature (T$_1$) of the temperature minimum (mesopause) (iii)
the temperature (T$_2$) and (iv) thickness ($\Delta$z$_2$) of the warm layer. In all the models, the base of the warm
layer was located 50 km above z$_1$, and the temperature above the top of the warm layer progressively returned 
to the nominal upper atmosphere temperature (68 K) within another 50 km. In practice, we tested three values of 
z$_1$ (300, 500, and 750 km) and three values of $\Delta$z$_2$ (100, 200 and 400 km), and searched for best fits 
of the HCN and CO lines in terms of T$_1$ and T$_2$. Note that these temperature profiles are purely empirical,
i.e. we do not examine the physical processes that could possibly drive them.
Given the sharp sensitivity of the HCN vapor pressure
with temperature, T$_2$ can be determined to within less than 1 K, while T$_1$, required to maintain a good fit of CO, is determined
somewhat more loosely. Results are summarized in Table 4 and Fig. 9 shows resulting fits along with
the temperature and HCN profiles in the $\Delta$z$_2$ = 200 km case.

\begin{table}
{\bf Table 4: Temperature profiles for non-supersaturated HCN profiles}

\vspace*{5mm}

\begin{tabular}{|c|c|c|c|}
\hline
z$_1$ (mesopause  &  $\Delta$z$_2$ (warm & Mesopause  & Warm layer \\
altitude, km) & layer thickness, km) & temp. T$_1$ (K) & temp. T$_2$ (K)  \\
\hline
300 &    100 &        57 &     98   \\
500 &    100 &        60 &     96 \\ 
700 &    100 &        50 &     94.5\\
300 &    200 &        56 &     96.5\\
500 &    200 &        60 &     94.5\\
700 &    200 &        50 &     93.5\\
300 &    400 &        56 &     93.5\\
500 &    400 &        60 &      92.5\\
700 &   400  &        50 &     92 \\
\hline
\end{tabular}
\end{table}

As is clear from Fig. 9, satisfactory solutions to the problem exist for z$_1$ = 500 and 700 km, and to a lesser
extent for z$_1$ = 300 km as well. The associated temperatures of the warm layer are in the 92-98 K range, and
the temperature minimum is $\sim$50-60 K. When z$_1$ = 300 km, the agreement deteriorates, as the rapid temperature
increase over 300-350 km, i.e. in a region well-probed by CO, tends to produce an undesired ``spike" in the CO line core.
Note that by construction, these solutions have significant HCN amounts in the upper atmosphere only down to 750, 550 and 350 km. In the first two cases, this is noticeably shallower than in one case considered before (uniform HCN mixing down to z$_0$ = 460 km and with q$_0$ = 4$\times$10$^{-5}$, see Fig. 6, blue dotted line); the difference is due to the inclusion, in the present exercise, of the lower atmosphere component, which by preferentially contributing to the satellite lines, modifies the required
distribution of HCN opacity in the upper atmosphere. 

We conclude that when allowance is made for the presence of low-altitude ($<$200 km) HCN, the CO and HCN data taken together
do not exclude the presence of a warm (92-98 K) upper layer located above $\sim$450 km, warm enough
that the HCN amounts indicated by the data do not require HCN supersaturation. However we will show below (Section 5.1.1.) that this
kind of profile is inconsistent with measurements from New Horizons/Alice.

\subsection{HC$^{15}$N and HC$_{3}$N upper limits}
As indicated in Section 3, the HC$^{15}$N(4-3) transition at 344.200 GHz was not detected with a 1$\sigma$ rms
noise level of 1.9 mJy per 0.488 MHz resolution. Using our best fit HCN model (light blue line from Fig. 6) and scaling
it by a vertically constant isotopic ratio factor, we determine that the non-detection implies a lower limit of the
HC$^{14}$N / HC$^{15}$N ratio equal to 125 at 2$\sigma$ confidence level (Fig. 10). Thus and interestingly, the data readily exclude
a Titan-like isotope ratio \citep[HC$^{14}$N / HC$^{15}$N $\sim$ 60;][]{vinatier07}. In contrast, the lower limit of 125
is consistent with the telluric ($^{14}$N/$^{15}$N $\sim$ 270), cometary ($^{14}$N/$^{15}$N $\sim$ 125-150) and primordial solar nebula ($^{14}$N/$^{15}$N $\sim$ 440, as measured in Jupiter, a meteoritic inclusion and the bulk Sun) isotope ratios 
\citep[for a review see][]{furi15}.

Translating the result into the bulk $^{14}$N/$^{15}$N ratio in Pluto's atmosphere, and ultimately into the $^{14}$N/$^{15}$N 
ratio in the building blocks that formed Pluto, is however a complex problem. 
On Titan, the high enrichment in HC$^{15}$N results from the combination
of a significant $^{15}$N enrichment in the bulk N$_2$ atmosphere  \citep[$^{14}$N/$^{15}$N  = 168, ] 
[]{niemann05,niemann10}  with a photolysis fractionation effect in the HCN formation, i.e. the 
self-shielding of $^{14}$N$_2$ from solar UV  \citep{liang07}. The bulk $^{14}$N/$^{15}$N ratio 
in Titan N$_2$ was initially interpreted by a large preferential escape of $^{14}$N$_2$ over the evolution of Titan \citep{niemann05}. This scenario was questioned by \citet{mandt14,mandt15}, who found that escape does not significantly fractionate Titan's N$_2$
and concluded that Titan's N$_2$ isotope ratio was primordial. Their result, however, was in turn questioned by \citet{johnson16}, who found that the escape rates used in \citet{mandt14} are not necessarily correct for fractionation.
Coming back to Pluto, it is difficult to evaluate whether the $^{14}$N/$^{15}$N ratio has significantly fractionated
over Pluto's history. The very small escape rates evidenced by New Horizons (Gladstone et al. 2016) tend to 
argue against this scenario, all the more that these measurements were acquired near perihelion. However, as
the mechanism that keeps the upper atmosphere cold is not yet identified (see discussion in Section
5.4) and could conceivably not operate further from the Sun, one cannot exclude stronger escape rates away from
perihelion. Furthermore, a number
of geological features evidenced by New Horizons, while not conclusive in themselves, may provide evidence 
for higher-pressure atmospheres (hence enhanced escape) in the past, a situation that might be expected at epochs 
when -- unlike in the present orbit -- orbital perihelion coincides with solstice, leading to ``extreme summers" 
(Stern et al. 2016). A second complication in interpreting Pluto's $^{14}$N/$^{15}$N ratio is that isotope fractionation
might also occur at sublimation and condensation so that the $^{14}$N/$^{15}$N in N$_2$ gas may not
necessarily reflect that in Pluto's ices. 
Fractionation at sublimation is observed for water $^{18}$O/$^{16}$O
and D/H in terrestrial snow and ice  \citep[][though fractionation factors are not available]{sokratov99} but to our knowledge the process is not documented for N$_2$. 
In any case, the first task would be to model out the photolytic fractionation effect on HCN. On Titan, the 
differential self-shielding of $^{14}$N$_2$ from solar UV leads to a maximum enhancement of the $^{14}$N$^{15}$N/ $^{14}$N$_2$
photolysis rate ratio by a factor 25 at 760 km altitude  \citep{liang07}. The equivalent pressure level (1.5$\times$10$^{-8}$ bar)
occurs at $\sim$340 km in Pluto's atmosphere. The HCN observations probe the entire atmosphere
from the surface to $\sim$900 km, i.e. presumably sample both regions of enhanced  HC$^{15}$N / HC$^{14}$N
ratio due to photolytic fractionation and regions of ``background" ratios. A detailed model would thus be needed for a proper interpretation of our upper limit 
in terms of the current $^{14}$N/$^{15}$N in Pluto's atmosphere.

 
Similarly, the HC$_{3}$N(38-37) and HC$_{3}$N(39-38) transitions at 345.609 and 354.697 GHz are not detected,
with 1$\sigma$ rms of 3.28 mJy per 0.141 MHz resolution for the former and 2.85 mJy per 0.282 MHz resolution
for the latter. Converting those into an upper limit of the HC$_{3}$N column density requires an assumption
on the species' vertical profile. Assuming here for simplicity that HC$_{3}$N has the same distribution as our
best fit HCN model (light blue line from Fig. 6), we find an upper limit of 1/10 for the HC$_{3}$N / HCN mixing
ratio, corresponding to a maximum HC$_{3}$N column density of 2$\times$10$^{13}$ cm$^{-2}$ (Fig. 10). 
Given the crude assumption of identical vertical distributions for HCN and HC$_3$N, this upper limit
should only be considered as typical, and should be re-derived with realistic photochemically-based vertical profiles
for HC$_3$N when those become available.

\section{Discussion}
\subsection{Thermal profile: comparison with New Horizons/Alice and ground-based stellar occultations measurements}
In the preceding, standard temperature profiles from New Horizons \citep{gladstone16} and ground-based occultations 
\citep{dias15} were only used as a priori guesses when retrieving thermal profiles from CO. Conversely, our
solution thermal profiles can be tested against these observational constraints.\\

\subsubsection{New Horizons/Alice} 
The primary observables relevant to thermal profile are the line-of-sight (LOS) column
densities of the major gases (N$_2$ and CH$_4$). These are the data that led to the establishment of the standard temperature
profile in \citet{gladstone16}. Using several of the previously discussed thermal profiles, we constructed a
diffusion model for the CH$_4$--N$_2$ mixture, using the formalism of \citet{yelle06} for two gases in comparable abundance.
Free parameters are the surface CH$_4$ mole fraction q$_{0,CH_4}$ and the vertical eddy diffusion coefficient K$_z$ profile.
We also specify an effective CH$_4$ loss rate of 1.2$\times$10$^{26}$ s$^{-1}$ at the top of the model, meant to represent
the combined effect of the actual CH$_4$ escape (with rate $\sim$5$\times$10$^{25}$ s$^{-1}$) and of the CH$_4$ photolysis -- which is not explicitly included in the model. As discussed by \citet{yelle08}, any other value of the escape rate can be equivalenty
handled by the use of another K$_z$ profile.
Fig. 11 compares the Alice LOS data (as shown in \citet{gladstone16}, and kindly provided to us by J. Kammer)
to models. LOS columns above 850 km for N$_2$ and 250 km for CH$_4$, which are the most reliable, are considered.
For the nominal thermal profile (black line), we find that a good fit of the CH$_4$ and N$_2$ LOS columns is achieved for 
q$_{0,CH_4}$ = 0.65 \% and K$_z$ = 1$\times$10$^{6}$ cm$^2$s$^{-1}$ above 200 km (the exact K$_z$ profile used is shown in the
inset of Fig. 10). q$_{0,CH_4}$ is generally consistent with the CH$_4$ abundance (0.3 - 0.6 \%)  measured in 2008 and 2012 by \citep{lellouch09,lellouch15a}. The solution parameters are slightly different but rather consistent from those given in
the SOM of \citeauthor{gladstone16} (q$_{0,CH_4}$ = 0.60-0.84 \%, K$_z$~=~(1.5---3)$\times$10$^{6}$ cm$^2$s$^{-1}$). We regard the fit in Fig. 11 as evidence that our nominal thermal profile is consistent with the Alice data.
In contrast, the alternate thermal profiles constructed previously to attempt to avoid HCN supersaturation (with a cold mesopause and a warm $\sim$95 K layer), are clearly at odds with the data, as they would produce an incorrect slope of the CH$_4$ LOS columns in some part of the atmosphere; additionally, except for the profile in which the warm layer occurs at the highest altitude,
these profiles lead to an overestimate of the N$_2$ LOS columns above 850 km altitude. We conclude that 
alternate solutions with an HCN profile at saturation, while not excluded from the sole point-of-view of the ALMA data,
are not viable, and we dismiss the associated thermal profiles.

\subsubsection{Stellar occultations}
As outlined previously, thermal profiles retrieved from CO robustly exhibit a 65-70 K temperature
at 250-400 km, i.e. somewhat colder than thermal profiles derived from stellar occultations \citep{dias15,sicardy16}, 
which have T = 81$\pm$6 K above 200 km. The difference (2$\sigma$) is barely significant, as occultation 
data bear little information on conditions above $\sim$250 km. However, our retrieved profiles are also colder 
than the occultation profiles by 4-10 K over 50-200 km, in which range they also show considerable variability
depending on the a priori used (see inset of Fig. 12). We calculated synthetic occultation
profiles for a variety of thermal profiles for direct comparison with observed
occultation curves; in practice, we used the high S/N 2012 July 18 occultation observed with VLT/NaCO.
Note that although our current thermal profile has a surface pressure of 12 $\mu$bar, we 
here rescaled the density profiles using p$_{surf}$ = 11.6 $\mu$bar, the value minimizing the residuals
of our nominal model with respect to observations.
Fig. 12 shows the comparison of the observed vs calculated fluxes
as a function of distance from Pluto's shadow center, for the five thermal profiles shown in inset. 
While, unsurprisingly, the ``cold Dias-Oliveira (DO15 + 69 K)" profile
matches the occultation data well, none of the other profiles provides a good representation of the data.
In particular, the New Horizons standard profile \citep{gladstone16}
is markedly inconsistent with the stellar occultation data, as can be best seen from the residuals in
the bottom of Fig. 12. Thus, although the published REX profiles from July 14, 2015 
(see Fig. 1 of \citet{gladstone16}) and the occultation profiles from \citet{dias15} both show T~$\sim$~110 K at 30-50 km, 
the mutual consistency between the two datasets will need to be assessed carefully\footnote{Note also that the REX-derived
surface pressure are 10$\pm$1 and 11$\pm$1 $\mu$bar, vs 11.9 -- 13.7 $\mu$bar based on
the June 29, 2015 stellar occultation \citep{sicardy16} and these small differences still  
remain to be reconciled.}. Thermal profiles retrieved from ALMA CO using either \citet{gladstone16}
or \citet{dias15} as a priori are also at odds with the occultation data. Finally, our nominal profile, while 
improving the fit compared to the three previous cases, still produces residual ``oscillations" (red curves) that indicate an inconsistency with the occultation data. Given these mismatches, the 11.6 $\mu$bar adopted in this exercise should
not be regarded as a new, improved value of p$_{surf}$ when compared with p$_{surf}$ = 11.15 $\mu$bar
as estimated for mid-2012 by \citet{dias15}.

Occultations probe near the terminators and at specific latitudes, while our observations are dayside and disk-integrated. 
However, it is unlikely that thermal profile differences arise from spatial and/or temporal (i.e., 2012 vs 2015) variations.
This is because radiative time constants in Pluto's
atmosphere are long \citep[estimated to be in the range 10--15 yr by][]{strobel96},
leading to small longitudinal and equator-to-pole temperature differences. Both \citet{toigo15}
and \citet{forget16} estimate that meridional variations in temperature are less than 1 K for
the New Horizons epoch, though, interestingly, \citeauthor{toigo15} (their Fig. 16) find that the pole-to-Equator gradient
in the middle atmosphere could have reached $\sim$10 K in the last few decades. We note however that the eight 
temperature profiles inverted from four lightcurves in 2012 and 2013 by \citet{dias15} show variability at the $\pm$5 K 
level in the first $\sim$200 km (although in no case is the stratopause colder than 109 K), which would not be 
expected in view of the above considerations, and if real might point to dynamical effects.


As discussed in Section 4.1.2 and Fig. 2, the colder temperatures over 50-200 km are required for an optimum fit of
the CO line core, so the inconsistency with the stellar occultation curves does not seem obvious to resolve. Admittedly,
the ``cold Dias-Oliveira" profile with upper atmosphere at 69 K would provide a better overall compromise when fitting
together the CO line and the VLT/NaCO occultation curve. However, the discrepancies outlined in Fig. 12 rather suggest that a
global fit of all the constraints relevant to Pluto's thermal structure (i.e., stellar occultations, REX, Alice solar 
occultation, CH$_4$ near-IR and ALMA CO spectroscopy) should eventually be performed. For now, 
while the well-marked mesosphere and the T = 70$\pm$2~K temperature at 300 km are well defined from our retrievals, 
the lower mesosphere and stratopause temperatures must be considered as more uncertain. As discussed above,
this does not impact our conclusions on the HCN distribution, but makes the interpretation of the HCN 
amounts in the lower atmosphere more ambiguous.  \\








\subsection{CO abundance}
We determine a CO mole fraction of 515 $\pm$ 40 ppm for a surface pressure p$_{surf}$ = 12 $\mu$bar,
rescaling as (12 $\mu$bar/p$_{surf}$)$^{1.9}$ for other values of p$_{surf}$. The associated 
surface-referred column density is 0.100$\pm$0.008 cm-am for p$_{surf}$ = 12 $\mu$bar, and scales 
as (12 $\mu$bar/p$_{surf}$)$^{0.9}$. These figures are much more accurate than, yet nicely consistent
with, results from Pluto's 2.3 $\mu$m spectrum \citep{lellouch11}. Specifically, these authors
inferred q$_{CO}$ = 500$^{+1000}_{-250}$ ppm, where most of the uncertainty was due to signal-to-noise
limitations (and to a much lower extent to the uncertain lower atmosphere thermal structure and surface 
pressure at that time). Based on Table 1 from \citet{lellouch11}, for p$_{surf}$ = 12 $\mu$bar, the best fit CO
mole fraction and column density from the 2.3 $\mu$m data would have been 530 ppm and 0.09 cm-am, respectively,
but still with a factor $\sim$2 uncertainty. Our new determination thus validates the ``strong evidence 
(6-$\sigma$)" for gaseous CO reported by  \citet{lellouch11} as being indeed a detection, while 
superseding it by far in accuracy. In contrast, the formal 4.5-$\sigma$ detection of the CO(2-1) line by \citet{bock01}, that was modelled in terms of a much higher 1.2 \% - 7 \% CO mixing ratio, must indeed have been
affected by galactic contamination, and as mentioned previously, was cautiously considered as such by the authors. 

Attempts to interpret the CO atmospheric abundance in terms of surface-atmosphere interactions were
presented by \citet{lellouch11}. Here the key parameter is the CO/N$_2$ mixing ratio in the surface ice. 
With the $\sim$0.5~\% CO/N$_2$ value originally inferred by \citet{owen93}, an atmospheric abundance of $\sim$500 ppm (0.05\%)
is consistent with an ideal solid solution in equilibrium with the atmosphere (i.e., Raoult's law). \citet{lellouch11} noted that this interpretation falls apart if updated CO/N$_2$ ice ratios of 0.08-0.2~\% 
\citep{doute99} were used. Instead, they proposed that the consistency, within error bars, of these latter values 
with their 0.025-0.15~\% CO gas mixing ratios favored the so-called ``detailed balancing model" proposed by \citet{trafton97}.
 An alternative explanation in terms of pure CO ice patches enriching the atmosphere in CO gas was also mentioned
but rejected on thermodynamical grounds. A recent analysis of new near-IR spectra \citep{merlin15} now suggests 
(disk-averaged) CO/N$_2$ ice ratios of 2500-5000 ppm, {\em not} consistent with our updated atmospheric mixing 
ratio of 515 ppm, challenging the detailed balancing explanation and favoring again the ideal solution interpretation. Even more recently, a post-New Horizons global climate model (GCM) for Pluto has been developed  \citep{forget16,bertrand16}.
In this model, which includes the N$_2$, CH$_4$ and CO cycles, volatiles sublimate according to Raoult's law and
their surface abundances. Regarding CO, model findings are that CO, which is almost as volatile as N$_2$, only condenses when N$_2$ ice is present at the surface, and never forms separate CO deposits. This is in agreement with the strongly localized distribution of CO ice in to Sputnik Planum \citep{grundy16}, where N$_2$ ice also appears concentrated (but not restricted to). In the GCM, the CO cycle is dominated by a condensation-sublimation cycle inside Sputnik Planum. Interestingly, the model, which uses a CO/N$_2$ ice ratio of 0.3~\% from \citet{merlin15}, predicts an atmospheric  CO/N$_2$ of 0.04~\%. Although this is not yet fully consistent with our value, it suggests that the essential physics is captured. Model predictions should be improved by using the CO ice mixing ratio in Sputnik Planum when the latter becomes available from New Horizons.

\subsection{HCN profile, supersaturation and photochemical models}
HCN column densities associated with our various line fits and retrievals span the range (1.2--2.0)$\times$10$^{14}$
cm$^{-2}$. Although the precise vertical distribution cannot be unambiguously determined from
lines of moderate optical thickness, the data support a bimodal HCN distribution, with a high-altitude ($>$500 km)
component with highly supersaturated (1.5-5)$\times$10$^{-5}$ mole fraction, and a near-stratopause component at 10$^{-7}$--10$^{-8}$.
In the best fit empirical model, the upper (resp. lower) atmosphere components have 5$\times$10$^{13}$ (resp.
1.5$\times$10$^{14}$) cm$^{-2}$ columns.

HCN is the prime nitrile to be expected in a N$_2$-CH$_4$
atmosphere and had been predicted for Pluto  \citep{lara97,summers97,krasno99,gladstone15}. Altogether, these models have envisaged a vast range of HCN mixing ratios 
and distributions, and it is a little difficult to track down the reasons for the differences, although
some must result from different temperature profile assumptions.
Nonetheless, almost all of these models appear to have severely overpredicted the
amounts of HCN in Pluto's atmosphere. For example, the HCN column density in \citet{krasno99} is quoted as 6$\times$10$^{16}$ cm$^{-2}$, a factor of 400 too large, and the HCN mixing ratio at 1000 km, $\sim$1$\times$10$^{-3}$ and 
$\sim$5$\times$10$^{-4}$ for models 1 and 2 in that paper, is 10--60 times too large. (Note also that their calculated HC$_3$N
column was 3.4$\times$10$^{16}$ cm$^{-2}$, more than three orders of magnitude larger than our upper limit).  Similarly, in 
the pre-New Horizons models by \citep{gladstone15}, most of the HCN profiles have a (3-5)$\times$10$^{-4}$
upper atmosphere mixing ratio. \citet{lara97}, who used a thermal profile with a marked temperature decrease above the stratopause (from Strobel et al. 1996) predicted a pile-up of HCN over 30--200 km with a typical $\sim$3$\times$10$^8$ cm$^{-3}$
number density in this range, a consequence of atmospheric condensation. Still, their HCN column was about 5$\times$10$^{15}$ cm$^{-2}$, a factor 30 too large. \citet{summers97}
also predicted HCN restricted to the lower atmosphere, with $\sim$10$^7$ cm$^{-3}$ over 50--150 km, 
implying a $\sim$10$^{14}$ cm$^{-2}$ column density, in rather good agreement with our measurement, but missing
the upper atmosphere component.

Once produced in the upper atmosphere, HCN must be transported downwards to a condensation sink, and be further
transported in solid form down to the warmer stratopause region where in principle it can re-evaporate. 
An order-of-magnitude estimate of the HCN production rate can be obtained by equating the
chemical column net production rate $P$ to the downward flux, i.e. $P$~=~a$_{HCN}$~/~$\tau$, where the
time constant $\tau$ is equal to 2$H^2$/$max$($K_z$,D$_{HCN-N_2}$). Here, a$_{HCN}$ is the upper atmosphere
component column density, $H$ is the atmospheric scale height, and $K_z$ and D$_{HCN-N_2}$ are the eddy
and molecular diffusion coefficients, respectively. 
Using a$_{HCN}$ = 5$\times$10$^{13}$ cm$^{-2}$, H = 65 km, $K_z$ =  10$^{6}$ cm$^{2}$s$^{-1}$ and 
D$_{HCN-N_2}$ = 1.5$\times$10$^{7}$ cm$^{2}$s$^{-1}$ at 450 km,
leads to a column net production rate  $P$ = 0.9$\times$10$^{7}$ cm$^{-2}$s$^{-1}$ at this altitude, i.e. $P$ = 1.7$\times$10$^{7}$ cm$^{-2}$s$^{-1}$ referred to the surface. This estimate should be compared to predictions based on photochemical
models. 

The most remarkable of our findings is the fact that Pluto's upper atmosphere, albeit at $\sim$70 K,
contains large HCN amounts, implying a huge supersaturation with respect to local conditions (e.g.
7-8 orders of magnitude at 800 km). Supersaturation of condensible species is known to occur in 
other planetary atmospheres. In the Earth's upper troposphere, enhancements of H$_2$O gas over
the gas-ice equilibrium are frequently observed, with relative humidity levels routinely of 120-150 \% and
(rarely) up to $\sim$250~\% \citep{gettelman06}. In the Martian atmosphere, supersaturation of water vapor is clearly observed, with saturation ratios up to 5-10 over 30-50 km altitude \citep{maltagliati11}, and large supersaturation of 
CO$_2$ in the upper atmosphere is indicated by the occasional existence of temperatures 
significantly below the CO$_2$ frost point over 90-120 km \citep{forget09}. Supersaturation is 
commonly assumed to result from the lack of condensation nuclei in clear atmospheres (or parts thereof),
restricting the role of heterogeneous nucleation and possibly leaving homogeneous nucleation as the
only mechanism for limiting the amount of condensibles in the gas phase. Nonetheless, supersaturation
ratios by orders of magnitude have not been observed before in a planetary atmosphere. The pressure 
at which HCN supersaturation occurs on Pluto ($\sim$2$\times$10$^{-8}$ mbar at 800 km altitude) is also
much lower than in the Earth ($\sim$200 mbar for H$_2$O in the upper troposphere) and Mars (typically $\sim$0.1 mbar for H$_2$O and
$\sim$10$^{-5}$ mbar for CO$_2$) cases. 

Although's Pluto's atmosphere is haze-rich, the haze in New Horizons images is detected up to $\sim$200 km only,
with a brightness scale height of 30 km over 100-200 km, and an estimated concentration of 0.8 particle cm$^{-3}$ 
at the surface \citep{stern15,gladstone16}. Assuming that the haze is well mixed with gas, the haze concentration 
would be $\sim$$10^{-6}$ particle cm$^{-3}$ at 800 km, where the pressure is about 2$\times$10$^{-11}$ bar.
These extremely small numbers may explain the strong inefficiency of neutral homogeneous and heterogeneous nucleations
in Pluto's upper atmosphere.
  
Development of a coupled photochemical-microphysical-ionospheric model for Pluto is in progress (Lavvas et al., in
prep.). The model is based on the \citet{gladstone16} standard temperature profile. Regarding condensation, as for now, the model includes homogeneous neutral nucleation as well as nucleation of HCN due to the presence of ions in Pluto's upper atmosphere (whose abundance is calculated self-consistently). A preliminary HCN profile resulting from the coupled model is shown in Fig. 7. Testing it against the ALMA observations indicates that a good agreement is obtained by uniformly dividing the calculated HCN 
mixing ratio in the model by 2. This rescaled profile has a HCN mole fraction of 4$\times$10$^{-5}$ at 800 km and a
column density of 1.6$\times$10$^{14}$ cm$^{-2}$, in full agreement with previous inferences from manual fitting
and line inversion. Although the physical model must still be fine-tuned and its sensitivity to thermal profile
assumptions studied, this agreement is encouraging and demonstrates that a huge degree of super-saturation is 
a realistic situation in Pluto's upper atmosphere. In the lower atmosphere, where haze condensation nuclei
are available and where the kinetics of the condensation process are more efficient, 
HCN tends to follow local saturation. Due to the temperature increase towards
the stratopause, the evaporation of ice particles re-enriches the atmosphere in HCN gas. In this respect,
an important observational constraint is that the HCN amount in the lower atmosphere is consistent with saturation
at 106 K, which is also the preferred stratopause temperature in our retrievals based on CO. If 
the stratopause is actually warmer (e.g. at 110 K, as suggested by stellar occultation and REX data),
then HCN must be {\em undersaturated} in this region \citep[by up to a factor $\sim$10, exceeding the uncertainty
on the vapor pressure expression from][]{fray09}. To account for this putative situation, 
Lavvas et al. (in prep.) propose that ice particles could be big enough to fall through the region of 
the temperature maximum faster than they can evaporate, or more likely that the abundance
of HCN (and several hydrocarbons) may be reduced by heterogeneous processes taking place at the
surface of the haze particles and that inhibit their evaporation.

Although photochemical models indicate that C$_2$H$_6$ is the main photochemically produced
species in Pluto's atmosphere, its vapor pressure (typically 10 orders of magnitude higher than
HCN) is such that it is expected to condense only a few kilometers above the surface.
Other species can condense at higher altitudes with significant mass flux. On the other hand, the 
New Horizons/Alice profiles indicate that several species (C$_2$H$_2$, C$_2$H$_4$, C$_2$H$_6$) are 
depleted in the gas phase below $\sim$300 km relative to expectations from pure homogeneous considerations \citep{gladstone16},
suggesting a heterogeneous loss to the haze. Thus, condensed HCN contributes significantly 
to the haze depending on the altitude, and HCN is for sure one of the first species condensing at high altitudes 
with a large mass flux.
In addition, the ultimate sedimentation of the haze particles should lead to 
the presence of HCN ice on the surface, although its detectability is uncertain due to the 
fact that seasonal exchanges of N$_2$ \citep[typically 10 g cm$^{-2}$ per Pluto year, e.g.][]{olkin15}
 and other volatiles (CH$_4$, CO) occur at a much faster rate, presumably burying HCN ice
under layers of more volatile species. We still note that a tentative detection of HCN ice has been reported
on Triton \citep{burgdorf10}, but so far was not confirmed.

\subsection{Implications for heating budget}
Even prior to their detection, CO and HCN have been identified as possible important species for Pluto's atmosphere heating budget.
\citet{lellouch94} initially suggested that in addition to the heating/cooling ``thermostat" role of
CH$_4$ first identified by \citet{yelle89}, cooling due to LTE emission from CO
rotational lines should be significant, given the then estimated abundance of CO (10$^{-4}$--10$^{-3}$). 
Using a much more extensive radiative-conductive model, \citet{strobel96} first suggested
the possibility of a mild negative temperature gradient (--(0.-0.1) K/ km)  in the 
1-2 $\mu$bar pressure range. These models were updated by \citet{zalucha11a,zalucha11b} and \citet{zhu14} to
include new observational constraints on Pluto's atmosphere composition (CH$_4$~=~5$\times$10$^{-3}$, CO~=~5$\times$10$^{-4}$) 
and near-surface structure \citep{lellouch09,lellouch11}.  Nonetheless, these updated models confirmed the essential 
features of the earlier \citet{strobel96} models, continuing to show only weak negative temperature gradients 
above the stratopause, typically a $\sim$5 K decrease over a 300 km range for a CO mixing ratio of 5$\times$10$^{-4}$. 

The existence and origin of a well-marked mesosphere with an upper atmosphere at $\sim$80 K was first discussed in \citet{dias15}, 
although a similar behaviour had in fact been observed in several earlier stellar occultation profiles.
Noting that the above N$_2$-CH$_4$-CO models were unable to explain the $\sim$30 K decrease they observed
over 30-180 km (i.e. a --0.2~K/km gradient) unless CO was about 40 times larger than observed (i.e. 200$\times$10$^{-4}$), 
these authors considered HCN cooling, by analogy with Titan's thermosphere, where radiation in the intense rotational 
lines of HCN equilibrates the solar UV heating rate \citep{yelle91}. Although \citet{dias15} did not design a complete radiative model, they estimated the amount of HCN needed to compete with rotational cooling due to a 200$\times$10$^{-4}$
CO mole fraction. They concluded that a uniform HCN profile at $\sim$5$\times$10$^{-5}$ was required to explain
their temperature profile, implying a strong supersaturation of HCN {\em throughout} the atmosphere.

With an upper atmosphere at $\sim$70 K, temperature profiles deduced from New Horizons/Alice and from CO in the present study confirm and exacerbate the magnitude of the mesospheric temperature decrease. Constraints on HCN
we infer in this study indicate that HCN cooling cannot be the sole cause for this gradient. Remarkably enough, the
well determined HCN mole fraction of 4$\times$10$^{-5}$ at 800 km compares well to the estimate from 
\citet{dias15}, but their estimate was based on a 80 K upper atmosphere temperature, vs 70 K
as now established. Furthermore, the HCN data strongly exclude that HCN is well mixed down to the surface
and thus cannot make up for the entire required cooling. The New Horizons/Alice hydrocarbon detections \citep{gladstone16}
indicate that C$_2$H$_2$ non-LTE vibrational cooling in its $\nu_5$ band at 13.7 $\mu$m may be important near the
warm stratopause at 30 km. However, detailed calculations to be published elsewhere (Strobel et al. in prep.) indicate
that C$_2$H$_2$ cooling falls off rapidly with increasing altitude, in conjunction with the pressure decrease
that causes this emission to become more non-LTE and the temperature decrease that shifts this emission even further
away from the Planck function maximum. Specifically, these calculations find that the C$_2$H$_2$ cooling rate falls below
that of CO below $\sim$230 km. At this altitude and up to $\sim$500 km, both our HCN line inversions
and the physical model from Lavvas et al. (in prep.) matching the ALMA data (i.e. rescaled by 1/2) indicate HCN mole fractions 
lower than 2$\times$10$^{-5}$, i.e. unable to produce enough cooling. Therefore, in addition to the HCN deficit in the upper atmosphere to explain a $\sim$70 K atmosphere, the most critical problem is the lack of an identified
cooling species in the 230-500 km region. To illustrate the magnitude of the problem, Fig. 13 shows calculated temperature profiles including cooling by CO, C$_2$H$_2$ (as constrained by the New Horizons solar occultation data) and HCN, using several HCN profiles, along with observational temperature profiles.
The comparison of model temperatures for the rescaled Lavvas et al. profile (dark blue) with a case with no HCN (light blue) demonstrates that HCN is a relatively minor cooling agent in Pluto's atmosphere, causing a mere reduction of the atmospheric temperatures by 5-10 K at most
above 400 km, and unable to explain temperatures below 80~K. Reaching the asymptotic 70~K temperature would require
an HCN profile (red) more similar to the pre-New Horizons model of \citet{gladstone15}, in excess of the observed
profile by a factor 20-200, and reaching 300-700 ppm over 400-800 km. Even in this case, the calculated temperatures over 100-300 km exceed those inferred
from New Horizons and from the present study, pointing to too small cooling rates. 

The above suggests 
that another radiatively active constituent (perhaps haze, or another as yet undetected gas) or non-radiative effects are required, but identifying those does not seem straightforward. 
Hazes in planetary atmospheres represent both sources and sinks of heat. For example, at Titan, \citet{tomasko08} show that with the current (Huygens-measured) thermal profile, haze contributes comparable cooling rates as gas (their Fig. 9). On the other hand, the warmer stratospheric 
temperatures at Titan compared to Saturn indicate that the net effect of Titan's haze is heating. Similarly at Saturn,
\citet{guerlet15} find that polar haze cools the upper atmosphere (p$<$0.1 mbar) by 5 K in winter, while in summer, its net effect is a 6 K warming of the polar middle stratosphere (p$<$30 mbar). For haze to be 
an overall cooling agent, it must be strongly scattering in the visible and absorbing in the far-IR. We also note
that in New Horizons images, haze is apparent only below 300 km, while the atmosphere's heat budget
requires a cooling agent up to $\sim$500 km. The presence of (undetected) haze above $\sim$450 km would also
tend to contradict the observed HCN supersaturation there, by providing condensation nuclei.

Adiabatic cooling due to hydrodynamic expansion of escaping gas is another possible heat sink, that provides 
a regulating mechanism to variations of solar UV heating. \citet{zhu14} found that for pre-New Horizons
estimates of the escape rates ($\sim$3.5$\times$10$^{27}$ N$_2$ s$^{-1}$, adiabatic cooling contributes a $\sim$ 10-30 K
temperature difference above 400 km (their Fig. 9a). The much lower escape rates now determined suggest
a more minor role for adiabatic cooling. Specificially, \citet{gladstone16} mention that adiabatic cooling becomes important
for a loss rate of 3$\times$10$^{27}$ amu s$^{-1}$ (= 2$\times$10$^{26}$ CH$_4$ s$^{-1}$), $\sim$4 times
larger than observed. In the standard New Horizons temperature model of \citet{gladstone16}, adiabatic cooling is $\sim$ 5 times smaller 
than EUV-FUV heating at 1400 km altitude and becomes progressively negligible at lower altitudes. Overall,
these possibilities do not seem too promising, and the heat budget of Pluto's atmosphere remains poorly understood.

\section{Summary}
Based on high signal-to-noise ALMA observations performed in June 2015, we report the first observation of CO in Pluto's atmosphere at sub-millimeter wavelengths, and the first detection of HCN. Based on radiative transfer fits including 
optimal inversion techniques, we find that the CO and HCN lines probe Pluto's atmosphere up to $\sim$450 km and $\sim$900 km
altitude, respectively, with a large contribution due to limb emission. We reach the following conclusions:
\begin{itemize}
\item The CO mole fraction in Pluto's atmosphere is 515$\pm$40 ppm for a 12~$\mu$bar surface pressure. While this
essentially confirms the amounts previously estimated from near-IR spectroscopy, the much higher accuracy 
demonstrates that the gas CO mole fraction is significantly lower than the CO:N$_2$ ratio in the ice
phase. This, along with recent climate model results including the N$_2$, CH$_4$ and CO cycles, 
favors the scenario of an ideal N$_2$:CO solid solution feeding both gases into the atmosphere, and more
complex interpretations such as the detailed-balancing model do not seem supported any longer.
\item The CO line profile gives clear evidence for a well-marked temperature decrease over $\sim$50-400 km altitude, 
with a best-determined temperature of 70$\pm$2 K at 300 km, somewhat lower than those previously estimated from stellar occultations
(81$\pm$6 K), and in agreement with recent inferences from New Horizons / Alice solar occultation data. The preferred temperature
profiles deduced from CO have a $\sim$106~K stratopause temperature, also colder than that deduced from stellar
occultation profiles ($\sim$110 K). A global fit of all relevant datasets should be eventually performed, with the goal of defining
the ``best" thermal profile for Pluto's atmosphere.
\item  With an HCN mole fraction $>$ 1.5$\times$10$^{-5}$ above 450 km and a best determined value of 4$\times$10$^{-5}$
near 800 km, Pluto's upper atmosphere is particularly HCN-rich. Although technically this could be the sign of a warm ($>$92~K) upper atmosphere layer, still consistent with constraints from CO, this situation would  violate the Alice CH$_4$ and N$_2$ measurements. Instead, the HCN mixing ratios imply a supersaturation of HCN in the upper atmosphere to a degree (7-8 orders of magnitude) hitherto unseen in planetary atmospheres, and probably related to the lack of condensation 
nuclei above the haze region. Optimum fit of the HCN line indicates that HCN is also present in the bottom $\sim$100 km of the atmosphere, with a 10$^{-8}$ - 10$^{-7}$ mole fraction which, depending on the precise stratopause temperature, could be at saturation or undersaturated by up to a factor $\sim$10.
\item The HCN column is (1.6$\pm$0.4)$\times$10$^{14}$ cm$^{-2}$, including $\sim$5$\times$10$^{13}$cm$^{-2}$
in the upper atmosphere ($>$ 450 km). This suggests a surface-referred net column production rate  of $\sim$2$\times$10$^{7}$ cm$^{-2}$s$^{-1}$.
\item Although HCN has been previously identified as a potential cooling agent in Pluto's atmosphere, the HCN 
amounts determined in this study appear unsufficient to explain the well-marked mesosphere and upper atmosphere 
temperature, so that the heat balance of Pluto's middle and upper atmosphere remains to be understood.
\item From the non-detection of HC$_3$N and HC$^{15}$N lines, we infer upper limits to the HC$_3$N column density ($<$
2$\times$10$^{13}$ cm$^{-2}$) and to the HC$^{15}$N / HC$^{14}$N isotopic ratio ($<$ 1/125). 

\end{itemize}

Given the strengths of the CO and HCN signals measured by ALMA, it will be possible, in the future, to obtain moderately 
spatially-resolved 
maps of their emission. Specific goals would include (i) the characterization of the thermal field over 50-400 km (ii)
the search for horizontal variations of the HCN content or of temperature in the upper atmosphere ($>$500 km) (iii) constraints on the upper atmosphere dynamics from (more challenging) direct wind measurements. The great sensitivity of ALMA also makes additional molecular searches promising. In the mid- and long-term, and along with other Earth-based facilities, ALMA will remain an indispensable asset to monitor the evolution of Pluto's atmosphere composition and thermal state as Pluto recedes from the Sun. \\

\vspace*{2cm}

{\bf Acknowledgements:}\\
 This paper is based on ALMA program 2013.1.00446.S. ALMA is a partnership of ESO (representing its member states), NSF (USA) and NINS (Japan), together with NRC (Canada), NSC and ASIAA (Taiwan), and KASI (Republic of Korea), in cooperation with the Republic of Chile. The Joint ALMA Observatory is operated by ESO, AUI/NRAO and NAOJ.
The National Radio Astronomy Observatory is a facility of the National Science Foundation operated under cooperative agreement by Associated Universities, Inc. Part of the research leading to these results has received funding from the
European Research Council under the European Community's H2020
(2014-2020/ERC Grant Agreement 669416 ``LUCKY STAR''). E.L. and P.L. acknowledge support
from the French ``Programme National de Plan\'etologie". A.S., D.F.S., H.W., L.Y. and X.Z. were supported by NASA under the New Horizons Project. We acknowledge useful discussions with N. Fray and B. Schmitt.

\newpage

\label{}

\bibliography{bibliography.bib}

\begin{thebibliography}{00}
\bibitem[Barnes(1993)]{barnes93}Barnes, P.J., 1993. A search for CO emission from the Pluto-Charon system. Astron. J. 106,
2540-2543.
\bibitem[Bertrand and Forget(2016)]{bertrand16} Bertrand, T., Forget, F., 2016. Climate physics explains observed glaciers and volatiles on Pluto. Nature, in press. 
\bibitem[Bockel\'ee-Morvan et al.(2001)]{bock01}Bockel\'ee-Morvan, D., Lellouch, E., Biver, N., et al., 2001. Search for CO gas in Pluto, Centaurs and Kuiper Belt objects at radio wavelengths. Astron. Astrophys. 377, 343-353.
\bibitem[Buie et al.(1997)]{buie97}Buie, M.W., Tholen, D.J., Wasserman, L.H.,1997. Separate lightcurves of Pluto and Charon, 
Icarus 125, 233-244. 
\bibitem[Burgdorf et al.(2010)]{burgdorf10}Burgdorf, M., Cruikshank, D.P., Dalle Ore, C.M., et al., 2010. A tentative identification of HCN ice on Triton. Astrophys. J. 718, L53-L57.
\bibitem[Butler(2012)]{butler12}Butler, B.J., 2012. Flux density models for Solar System bodies in CASA. ALMA Memo 594, November 2012.
\bibitem[Butler et al.(2015)]{butler15}Butler, B., Gurwell, M., Lellouch, E., et al., 2015. Long wavelength observations of thermal emission from Pluto and Charon with ALMA. 47, id.210.04.
\bibitem[Conrath et al.(1998)]{conrath98}Conrath, B.J., Gierasch, P.J., Ustinov, E.A., 1998. Thermal structure
and para hydrogen fraction on the outer planets from Voyager IRIS measurements. Icarus 135, 501–517.

\bibitem[Cornwell and Fomalont(1999)]{cornwell99}Cornwell, T., Fomalont, E.B., 1999. Self-calibration. In G. B. Taylor, C. L. Carilli, and R. A. Perley, editors, Synthesis Imaging in Radio Astronomy II, Astr. Soc. Pacific Conf. Series, 180, 187-199.
\bibitem[Cordiner et al.(2015a)]{cordiner15a}Cordiner, M.A.,Palmer, M.Y., Nixon, C.A., et al., 2015a. Ethyl cyanide on Titan: spectroscopic detection and mapping using ALMA. Astrophys. J. 800, article id. L14, 7 pp.
\bibitem[Cordiner et al.(2015b)]{cordiner15b}Cordiner, M.A.,Palmer, M.Y., Nixon, C.A., et al., 2015b. ALMA spectroscopy of Titan's atmosphere: first detections of vinyl cyanide and acetonitrile isotopologues. Bull. Amer. Astron. Soc. 47, id.205.03.
\bibitem[Dias-Oliveira et al.(2015)]{dias15}Dias-Oliveira, A., Sicardy, B., Lellouch, E., et al., 2015. 
Pluto's atmosphere from stellar occultations in 2012 and 2013. Astrophys. J. 811, article id. 53, 20 pp. 
\bibitem[Dumouchel et al.(2010)]{dumouchel10}Dumouchel, F., Faure, A., Lique, F., 2010.	
The rotational excitation of HCN and HNC by He: temperature dependence of the collisional rate coefficients. MNRAS
406, 2488-2492.
\bibitem[Dout\'e et al.(1999)]{doute99}Dout\'e, S., Schmitt, B., Quirico, E., Owen, T.C., Cruikshank, D.P., de Bergh, C., Geballe, T.R., Roush, T.L., 1999. Evidence for methane segregation at the surface of Pluto. Icarus 142, 421-444. 
\bibitem[Elliot et al.(2003)]{elliot03}Elliot, J.L., Ates, A., Babcock, B.A., et al., 2003. The recent expansion of Pluto's atmosphere. Nature 424, 165-168. 
\bibitem[Forget et al.(2009)]{forget09}Forget, F., Montmessin, F., Bertaux, J.-L., et al., 2009. Density and temperatures of the upper Martian atmosphere measured by stellar occultations with Mars Express SPICAM. J. Geophys. Res. 114, E1, CiteID E01004.
\bibitem[Forget et al.(2016)]{forget16}Forget, F., Bertrand, T., Vangvitchith, M., et al., 2016. A post-New Horizons global climate model of Pluto including the N$_2$ , CH$_4$ and CO cycles. Icarus, submitted.
\bibitem[Fouchet et al.(2016)]{fouchet16}Fouchet, T., Greathouse, T., Spiga, A., Fletcher, L, Guerlet, S., 2016. Stratospheric aftermath of the 2010 storm on Saturn as observed by the TEXES instrument. I. Temperature structure. Icarus, 277, 196-214.
\bibitem[Fray and Schmitt(2009)]{fray09}Fray, N., and Schmitt, B., 2009. Sublimation of ices of astrophysical interest: A bibliographic review. Planet. Space Sci. 57, 2053-2080.
\bibitem[F\"uri and Marty(2015)]{furi15}F\"uri, E., Marty, B., 2015. Nitrogen isotope variations in the Solar System. Nature Geoscience, 8, 515-522.
\bibitem[Gettelman et al.(2006)]{gettelman06}Gettelman, A., Fetzer, E.J., Eldering, A., Irion, F.W., 2006.	
The global distribution of supersaturation in the upper troposphere from the Atmospheric Infrared Sounder. J. Climate 19, 6089-6103.
\bibitem[Gladstone et al.(2016)]{gladstone16}Gladstone, G.R., Stern, S.A., Ennico, K., et al., 2016. The atmosphere of Pluto as observed by New Horizons. Science 351, aad8866.
\bibitem[Greaves et al. (2011)]{greaves11}Greaves, J.S., Helling, C., Friberg, P., 2011. Discovery of carbon monoxide in the upper atmosphere of Pluto. MNRAS 414, L36-L40.
\bibitem[Guerlet et al. (2015)]{guerlet15}Guerlet, S., Fouchet, T. Vinatier, S., et al., 2015.	Stratospheric benzene and hydrocarbon aerosols detected in Saturn's auroral regions. Astron. Astrophys., 580, id.A89, 9 pp.
\bibitem[Gladstone et al.(2015)]{gladstone15}Gladstone, G.R.; Yung, Y.L., Wong, M.L. 2015, Pluto atmosphere photochemical models for New Horizons, 46th LPSC Conference, 46.3008G.
\bibitem[Grundy et al.(2013)]{grundy13}Grundy, W.M., Olkin, C.B., Young, L.A., Buie, M.W., Young, E.F., 2013. Near-infrared
  spectral monitoring of Pluto's ices: Spatial distribution and secular evolution. Icarus 223, 710-721. 
\bibitem[Grundy et al.(2016)]{grundy16}Grundy, W.M., Binzel, R.P., Buratti, B.J., et al., 2016. Surface compositions across Pluto and Charon. Science 351, aad9189.
\bibitem[Gurwell et al.(2014)]{gurwell14}Gurwell, M.A., Butler, B.J., Moullet, A., 2014. Atmospheric CO on Pluto: limits from millimeter-wave spectroscopy. Bull. Amer. Astron. Soc. 46, id.401.05.
\bibitem[Gurwell et al.(2015)]{gurwell15}Gurwell, M., Lellouch, E., Butler, B., et al., 2015. Detection of atmospheric CO on Pluto with ALMA. Bull. Amer. Astron. Soc. 47, id.105.06.
\bibitem[Hansen et al.(2015)]{hansen15}Hansen, C.J., Paige, D.A., Young, L.A., 2015. Pluto's climate modeled with new observational constraints. Icarus 246, 183-191.
\bibitem[Johnson et al.(2016)]{johnson16}Johnson, R.E., Tucker, O.J., Volkov, A.N., 2016.  Evolution of an early Titan atmosphere. Icarus, 271, 202-206.
\bibitem[Koshelev and Markov(2009)]{koshelev09}Koshelev, M.A., Markov, V.N., 2009. Broadening of the J=3-2 spectral line of carbon monoxide by pressure of CO, N$_2$ and O$_2$. J. Quant. Spectrosc. Radiat. Transf. 110, 526-527.
\bibitem[Krasnopolsky and Cruikshank(1999)]{krasno99}Krasnopolsky, V.A, Cruikshank, D.P., 1999. Photochemistry of Pluto's atmosphere and ionosphere near perihelion. J. Geophys. Res. 104, E9, 21979-21996.
\bibitem[van Langevelde and Cotton(1990)]{vanlange90}van Langevelde, H.J., and Cotton, W.D, 1990. Visibility-based continuum subtraction in spectral line observations with radio synthesis telescopes. Astron. Astrophys. 239, L5–L8.
\bibitem[Lara et al.(1997)]{lara97}Lara, L.M., Ip, W.-H., Rodrigo, R., 1997.	Photochemical models of Pluto's atmosphere.
Icarus 130, 16-35.
\bibitem[Lellouch (1994))]{lellouch94}Lellouch, E., 1994. The thermal structure of Pluto's atmosphere: clear vs hazy models.
Icarus 108, 255-264.
\bibitem[Lellouch et al.(2009)]{lellouch09}Lellouch, E., et al., 2009. Pluto's lower atmosphere structure and methane abundance from
high-resolution spectroscopy and stellar occultations. Astron. Astrophys. 495, L17-L21.
\bibitem[Lellouch et al.(2011)]{lellouch11}Lellouch, E., de Bergh, C., Sicardy, B., K\"aufl, H.U., Smette, A., 2011. 
High resolution spectroscopy of Pluto's atmosphere: detection of the 2.3 $\mu$m CH$_4$ bands and evidence for
carbon monoxide. Astron. Astrophys. 530, L4.
\bibitem[Lellouch et al.(2015a)]{lellouch15a}Lellouch, E., de Bergh, C., Sicardy, B., et al., 2015a, Exploring the spatial, temporal, and vertical distribution of methane in Pluto's atmosphere. Icarus 246, 268-278.
\bibitem[Lellouch et al.(2015b)]{lellouch15b}Lellouch, E., Gurwell, M., Butler, B. et al., 2015b. (134340) Pluto. IAUC 9273. 
\bibitem[Lellouch et al.(2015c)]{lellouch15c}Lellouch, E., Gurwell, M., Butler, B. et al., 2015c. Detection of HCN in Pluto's atmosphere. Bull. Amer. Astron. Soc. 47, id.105.07.
\bibitem[Liang et al.(2007)]{liang07}Liang, M.-C., Heays, A.N., Lewis, B.R., Gibson, S.T., Yung, Y.L., 2007.
Source of nitrogen isotope anomaly in HCN in the atmosphere of Titan. Astrophys. J. 664, L115-L118.
\bibitem[Maltagliati et al.(2011)]{maltagliati11}Maltagliati, L., Montmessin, F., Fedorova, A., Korablev, O., 
Forget, F., Bertaux, J.-L., 2011, Evidence of water vapor in excess of saturation in the atmosphere of Mars. Science 333, 
1868-1871.
\bibitem[Mandt et al.(2014)]{mandt14}Mandt, K., Mousis, O., Lunine, J., Gautier, D., 2014.	
Protosolar ammonia as the unique source of Titan's nitrogen. Astrophys. J. 788, id. L24, 5 pp.
\bibitem[Mandt et al.(2015)]{mandt15}Mandt, K., Mousis, O., Chassefi\`ere, E., 2015. Comparative planetology of the history of nitrogen isotopes in the atmospheres of Titan and Mars. Icarus,  254, 259--261.
\bibitem[Merlin(2015)]{merlin15}Merlin, F., 2015. New constraints on the surface of Pluto. Astron. Astrophys. 582, id.A39, 9 pp.
\bibitem[Muders et al.(2014)]{muders14}Muders, D., Wyrowski,  F., Lightfoot, J., et al., 2014. The ALMA Pipeline. In N. Manset and P. Forshay, editors, Astronomical Data Analysis Software and Systems XXIII, May 2014.
, 2014. The ALMA Pipeline. In N. Manset and P. Forshay, editors, Astronomical Data Analysis Software and Systems XXIII, May 2014.
\bibitem[M\"uller et al.(2011)]{muller01}M\"uller, H.S.P., Thorwirth, S., Roth, D.A., Winnewisser, G., 2011. The Cologne Database for Molecular Spectroscopy, CDMS. Astron. Astrophys., 370, L49-L52.
\bibitem[Niemann et al.(2005)]{niemann05}Niemann, H.B., Atreya, S.K., Bauer, S.J., et al., 2005. The abundances of constituents of Titan's atmosphere from the GCMS instrument on the Huygens probe. Nature 438, 779-784.
\bibitem[Niemann et al.(2010)]{niemann10}Niemann, H.B., Atreya, S.K., Demick, J.E., et al., 2010. Composition of Titan’s lower atmosphere and simple surface volatiles as measured by the Cassini‐Huygens probe gas
chromatograph mass spectrometer experiment. J. Geophys. Res.,  115, E12006, doi:10.1029/2010JE003659.
\bibitem[Olkin et al.(2015)]{olkin15}Olkin, C.B., Young, L.A., Borncamp, D., et al., 2015. Evidence that Pluto's atmosphere does not collapse from occultations including the 2013 May 04 event. Icarus 246, 220-225.
\bibitem[Owen et al.(1993)]{owen93}Owen, T.C., Roush, T.L., Cruikshank, D.P., et al., 1993. Surface ices and atmospheric
composition of Pluto. Science 261, 745-748.
\bibitem[Priem et al.(2000)]{priem00}Priem, D., Rohart, F., Colmont, J.-M., Wlodarczak, G., Bouanich, J.-P., 2000.	
	Lineshape study of the J=3-2 rotational transition of CO perturbed by N$_2$ and O$_2$. J. Mol. Spectro. 517, 435-454.
\bibitem[Rezac et al.(2013)]{rezac13}Rezac, L., Kutepov, A.A., Faure, A., Hartogh, P., Feofilov, A.G., 2013.	
	Rotational non-LTE in HCN in the thermosphere of Titan: implications for the radiative cooling. Astron. Astrophys. 555,
A122.
\bibitem[Rodgers et al.(2000)]{rodgers00}Rodgers, C.D., et al., 2000. Inverse methods for atmospheric sounding: theory and practice. Vol. 2. World Scientific, Singapore.
\bibitem[Schmitt et al.(2016)]{schmitt16}Schmitt, B., et al., 2016. Physical state and distribution of materials at the surface
of Pluto from New Horizons LEISA imaging spectrometer. Submitted for publication.
\bibitem[Sicardy et al.(2003)]{sicardy03}Sicardy, B., Widemann, T., Lellouch, E., et al., 2003. Large changes in Pluto's atmosphere as revealed by recent stellar occultations. Nature 424, 168-170.
\bibitem[Sicardy et al.(2016)]{sicardy16}Sicardy, B., Talbot, J., Meza, E., et al., 2016. Pluto's atmosphere from the 2015 June 29 ground-based stellar occultation at the time of the New Horizons flyby. Ap. J. 819, L38. 
\bibitem[Sokratov and Golubev(1999)]{sokratov99}Sokratov, S.A., and Golubev, V.N., 1999. Snow isotopic content change by sublimation. J. Glaciology, 55, 823-828.
\bibitem[Stern et al.(2015)]{stern15}Stern, S.A., Bagenal, F., Ennico, K., et al., 2015. The Pluto system: Initial results from its exploration by New Horizons. Science 350, aad1815.
\bibitem[Strobel et al.(1996)]{strobel96}Strobel, D.F., Zhu, X., Summers, M.E., Stevens, M.H., 1996.	
On the vertical thermal structure of Pluto's atmosphere. Icarus 120, 266-289.
\bibitem[Summers et al.(1997)]{summers97}Summers, M.E., Strobel, D.F., Gladstone, G.R., 1997. Chemical models of Pluto's atmosphere In: Stern, S.A., Tholen, D.J. (Ed.), Pluto and Charon, p. 391.
\bibitem[Thompson et al.(2001)]{thompson01}Thompson, A.R., Moran, J.M., and Swenson, G.W., 2001. Interferometry and Synthesis in Radio Astronomy, 2nd Edition. Wiley-Interscience, New York, New York.
\bibitem[Toigo et al.(2015)]{toigo15}Toigo, A.D., French, R.G., Gierasch, P.J., et al., 2015. General circulation models of the dynamics of Pluto's volatile transport on the eve of the New Horizons encounter 
Icarus 254, 306-323.
\bibitem[Tomasko et al.(2008)]{tomasko08}Tomasko, M.G., B\'ezard, B., Doose, L., et al., 2008. Heat balance in Titan's atmosphere.
Planet. Space Sci., 56, 648-659.
\bibitem[Trafton et al.(1997)]{trafton97}Trafton, L.M., Hunten, D.M., Zahnle, K.J., McNutt Jr., R.L., 1997. Escape processes at Pluto
and Charon. In: Stern, S.A., Tholen, D.J. (Eds.), Pluto and Charon. Univ. of Arizona Press, Tucson, pp. 475–522.
\bibitem[Vinatier et al.(2007)]{vinatier07}Vinatier, S., B\'ezard, B., Fouchet, T., et al., 2007.  Vertical abundance profiles of hydrocarbons in Titan's atmosphere at 15$^{\circ}$S and 80$^{\circ}$N retrieved from Cassini/CIRS spectra. Icarus 188, 120-138.
\bibitem[Yang et al.(2008)]{yang08}Yang, C., Buldyreva, J., Gordon, I.E., et al., 2008. Oxygen, nitrogen and air broadening of HCN spectral lines at terahertz frequencies. J. Quant. Spectrosc. Radiat. Transf. 109, 2857-2868.
\bibitem[Yelle(1991))]{yelle91}Yelle, R.V., 1991. Non-LTE models of Titan's upper atmosphere. Astrophys. J. 383, 380-400.
\bibitem[Yelle et al.(2006)]{yelle06}Yelle, R.V., Borggren, N., de la Haye, V., et al., 2006.
The vertical structure of Titan's upper atmosphere from Cassini Ion Neutral Mass Spectrometer measurements, Icarus 182, 567--576.
\bibitem[Yelle et al.(2008)]{yelle08}Yelle, R.V., Cui, J., M\"uller-Wodarg, I.C.F., 2008. Methane escape from Titan's atmosphere.
J. Geophys. Res. 113, E10, CiteID E10003.
\bibitem[Yelle \& Lunine(1989)]{yelle89}Yelle, R.V., Lunine, J.I., 1989. Evidence for a molecule heavier than methane in the
atmosphere of Pluto. Nature 339, 288-290.
\bibitem[Young(2013)]{young13}Young, L.A., 2013. Pluto's seasons: new predictions for New Horizons. Ap. J. Lett. 766, L22
(6pp).
\bibitem[Zalucha et al.(2011a)]{zalucha11a}Zalucha, A., Gulbis, A.A.S., Zhu, X., Strobel, D.F., Elliot, J.L., 2011a. An analysis of Pluto occultation light curves using an atmospheric radiative-conductive model. Icarus 211, 804-818.
\bibitem[Zalucha et al.(2011b)]{zalucha11b}Zalucha, X., Zhu, A.M., Gulbis, A.A.S., Strobel, D.F., Elliot, J.L., 2011b. An investigation of Pluto's troposphere using stellar occultation light curves and an atmospheric radiative-conductive-convective model. Icarus 214, 685-700.
\bibitem[Zhu et al.(2014)]{zhu14}Zhu, X., Strobel, D.F. and Erwin, J.T. 2014. The density and thermal structure of Pluto's
atmosphere and associated escape processes and rates. Icarus 228, 301-314.
\bibitem[Young et al.(1997)]{young97}Young, L.A., Elliot, J.L., Tokunaga, A., de Bergh, C., Owen, T., 1997. Detection of gaseous
methane on Pluto. Icarus 127, 258-262.
\bibitem[Young et al.(2001)]{young01}Young, L. A., Cook, J.C., Yelle, R. V., Young, E. F., 2001. Upper limits on gaseous CO at
Pluto and Triton from high-resolution near-IR spectroscopy. Icarus 153, 148-156.
\bibitem[Zangari(2015)]{zangari15}Zangari, A., 2015. A meta-analysis of coordinate systems and bibliography of their use on Pluto from Charon's discovery to the present day. Icarus 246, 93-145. 
\end{thebibliography}




\newpage

\begin{center}
{\large Figure captions} \\
\end{center}

\vspace*{.5 cm}

Fig. 1. Overview of CO(3-2) (top) and HCN(4-3) (bottom) line observations on the two observing days. UT date
at mid-integration is indicated.  For HCN, the hyperfine
structure is detected on both days from weak emission at -1.6 and +2.0 MHz from the main line. The continuum
has been subtracted. \\

Fig. 2. (Left) Fits of the average (June 12 + 13, 2015) CO(3-2) line using several temperature profiles
and a surface pressure of 12 $\mu$bar. Green and dashed : model calculations with the \citet{dias15} profile and CO = 500 ppm. Blue: best fit with a modified temperature profile (shown in blue in right panel) having T = 69 K above 300 km.
Red: best fit after simultaneous inversion of CO mole fraction and thermal profile, using the \citet{dias15} profile
as a priori. (Right): thermal profiles. \\

Fig. 3. Set of a priori (dashed) and retrieved (solid) temperature profiles, using a surface pressure of 12 $\mu$bar. (Left). CO is fixed at 500 ppm. (Right). CO is retrieved simultaneously (see Table 3 for results). Light blue, black and dark blue curves
refer to isothermal a priori profiles at 40, 70 and 100 K, respectively. Red and green curves refer to the temperature profiles 
of \citet{dias15} and \citet{gladstone16} as a priori.  \\

Fig 4. Information content of the CO(3-2) line. Averaging kernels at 0 (purple, solid line), 100 (dark blue, dotted), 200
(light-blue, short-dashed), 300 (green, dashed-dot), 500 (yellow, dashed-three dots) and 700 km (red, long-dashed) are shown.
They are calculated for the CO, T(z) solution achieved using the \citet{dias15} profile as a priori. 
While kernels at 0, 100, 200 and 300 km peak near the corresponding altitude, kernels at 500 and 700 km peak at 300-350 km,
indicating that temperatures at and above 500 km are not determined from the measurements themselves, but rather from the a priori profile and the correlation with the retrieved temperatures at lower altitudes.
The solid black line (upper x scale) shows the total kernel as a function of altitude.  \\ 

Fig. 5. Effect of surface pressure and calibration uncertainty on retrieved thermal profiles. For nominal
calibration: p = 12 $\mu$bar (red, solid line), p = 10 $\mu$bar (green, long-dashed), p = 14 $\mu$bar (blue, short-dashed). 
For p = 12 $\mu$bar, the effect of multiplying data by factors of 1.05 (light blue, dashed-dotted) and 0.95 (pink, dots)
is also shown. All profiles are retrieved using the ``mixed" a priori profile (see text) shown in black. \\

Fig. 6. Manual fits of the HCN (4-3) average (June 12 + 13, 2015) line, using the nominal temperature profile (shown in red in Fig. 5 and here in black in the top right panel), a surface pressure of 12 $\mu$bar, and a variety of HCN distributions, as shown in the bottom right panel. The yellow curve in the bottom right panel is the best fit two-component distribution obtained
if the ``cold Dias-Oliveira" temperature profile (yellow curve in upper right panel) is used. The yellow and light blue curves
are indistiguishable above 450 km.\\

Fig. 7. Vertical profiles of HCN (solid lines) retrieved from the HCN (4-3) line using a different set of a priori HCN profiles (dashed lines with same color code as the associated solid lines). The thin dotted line is the saturation profile for the adopted nominal thermal profile.
The solid black line marked ``Physical" is the HCN vertical profile in the preliminary model of Lavvas et al. (see
text). Dashed line: same profile, divided by 2, allowing a match of the HCN line. \\

Fig. 8. Information content of the HCN(4-3) line. Averaging kernels at 200 (purple, solid line), 400 (dark blue, dotted), 600
(light-blue, short-dashed), 800 (green, dashed-dot), 1000 (yellow, dashed-three dots) and 1200 km (red, long-dashed) are shown.
The solid black line (upper x scale) shows the total kernel as a function of altitude. These kernels are calculated
for the HCN solution profile returned using a constant HCN = 1$\times$10$^{-5}$ mole fraction as a priori (red curve in Fig. 7). \\

Fig. 9. Simultaneous fit of CO and HCN line with thermal profiles exhibiting a warm layer and assuming local saturation
of HCN at all levels. In the examples shown, the warm layers start 50 km above the mesopause and are 200 km thick.
Red, green, and blue curves correspond to mesopause altitudes of 700, 500, and 300 km respectively. In the top-left panel,
the 700 and 500 km mesopause cases lead to indistinguishable CO synthetic lines, while the 300 km
case tends to produce an undesired spike in the core of the CO line. \\

Fig. 10. Upper limits on HC$^{15}$N and HC$_3$N. Top: Average (June 12 + 13, 2015) Pluto spectrum in the region of the HC$^{15}$N(4-3) line at 344.200 GHz, compared with models using our preferred HCN distribution (light blue solid line in Fig. 6) and several values of the HC$^{14}$N / HC$^{15}$N ratio. A lower limit of 125 is indicated by the non-detection of HC$^{15}$N. Middle and bottom panels:
Average Pluto spectrum in the region of the HC$_{3}$N(39-38) and HC$_{3}$N(38-37) lines at 354.697 GHz and 345.609 GHz,
compared with a model using the same HCN distribution and a HC$_{3}$N / HCN = 1/10 ratio. \\

Fig. 11. New Horizons / Alice line-of-sight (LOS) column densities for CH$_4$ and N$_2$ (blue dots, ingress
and egress mixed). LOS columns above 850 km for N$_2$ and 250 km for CH$_4$, which are the most reliable, are considered.
Data are compared to predictions from a diffusion model with a surface CH$_4$ mole fraction of 0.65 \%, and the eddy K$_z$ 
and the four temperature profiles shown in the inset. While the nominal profile from this work (black line)
matches the Alice data, profiles exhibiting a strong temperature inversion and a $\sim$95 K layer
in the upper atmosphere (red, green, blue) are at odds with the CH$_4$ (and N$_2$, in general) LOS columns in
some parts of the atmosphere. \\

Fig. 12. Stellar occultation data from July 18, 2012 (VLT/NaCO) (light blue points) are compared with models using the temperature profiles
shown in the inset. These include: the ``cold Dias-Oliveira" profile (DO15+69 K; black), the standard New Horizons profiles (blue)
and three profiles derived from CO line fitting, using different a priori profiles. In the inset, dots indicate the New Horizons/REX measurements (red: immersion, black: emersion).
{\em Upper plots}. Data (normalized stellar flux, ingress and egress folded on each other) and models are plotted vs distance to shadow center.
{\em Lower plots}. Light blue: residuals of the July 18, 2012 data with respect to the best model
from \citet{dias15}. Other curves: residuals of the individual models (with same color coding) 
with respect to the best model from \citet{dias15}. All the residuals have been shifted by -0.1 for better clarity.\\

Fig. 13. Observational thermal profiles from \citet[]{dias15} (yellow),  \citet[][green line]{gladstone16}
and this work (nominal, black line), are compared to profiles calculated from a radiative-conductive model including CH$_4$ heating/cooling, CO rotational cooling, C$_2$H$_2$ vibrational cooling, and HCN rotational cooling for different HCN distributions (dashed lines, upper x scale). Light blue: No HCN. Dark blue: rescaled HCN distribution from Lavvas et al. (in prep.), also shown in Fig. 7, and that allows a match of the HCN line ALMA data. The associated thermal profile never gets colder
than $\sim$78 K. Red: a much enhanced HCN distribution, that permits
an approximate fit of the observed thermal profiles, but at the expense of 20-200 times too high HCN abundances.
Calculations from Strobel et al. (in prep.).
\pagestyle{empty}

\newpage

\epsfig{file=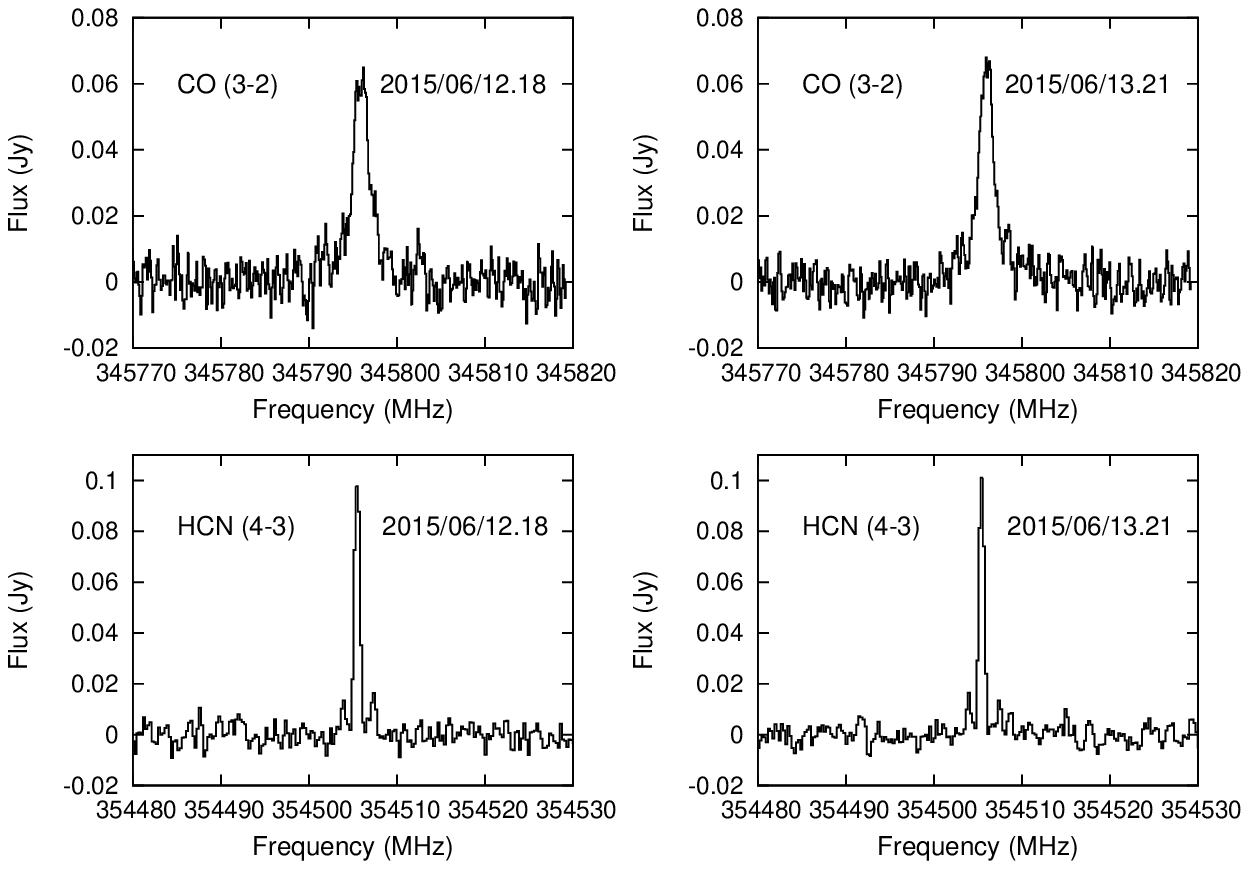,angle=0,width=15cm}

\vspace*{.5cm}
Fig. 1

\newpage
\epsfig{file=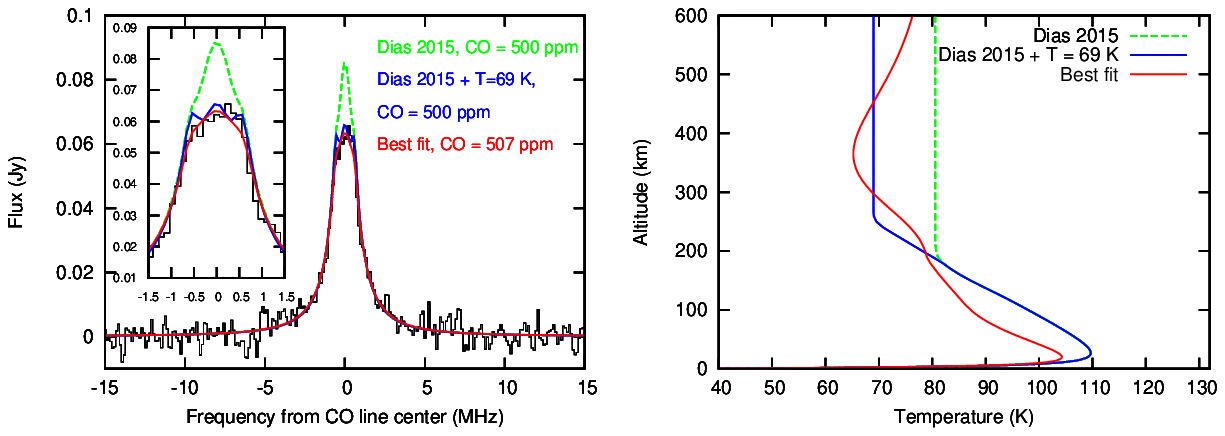,angle=-90,width=15cm}

\vspace*{.5cm}
Fig. 2

\newpage
\epsfig{file=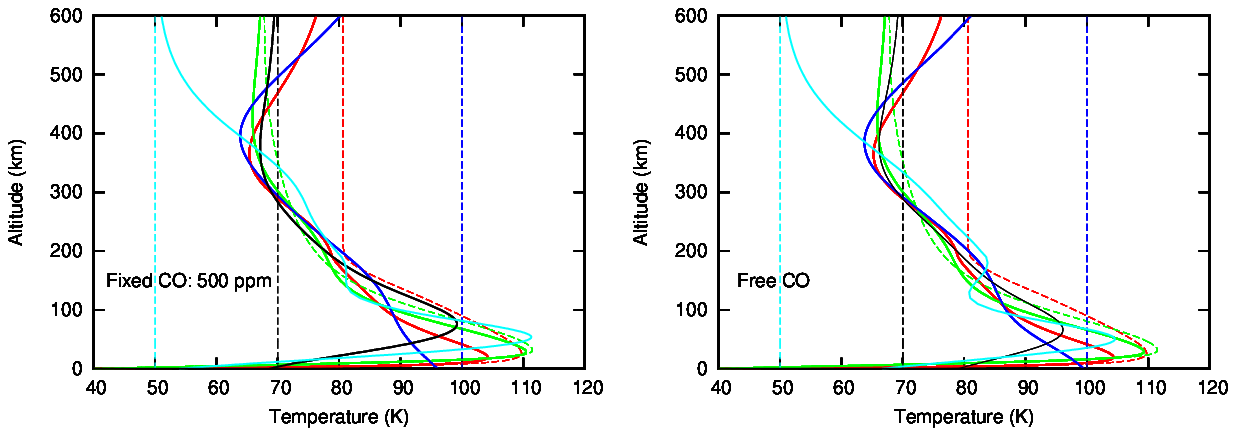,angle=-90,width=15cm}

\vspace*{.5cm}
Fig. 3

\newpage
\epsfig{file=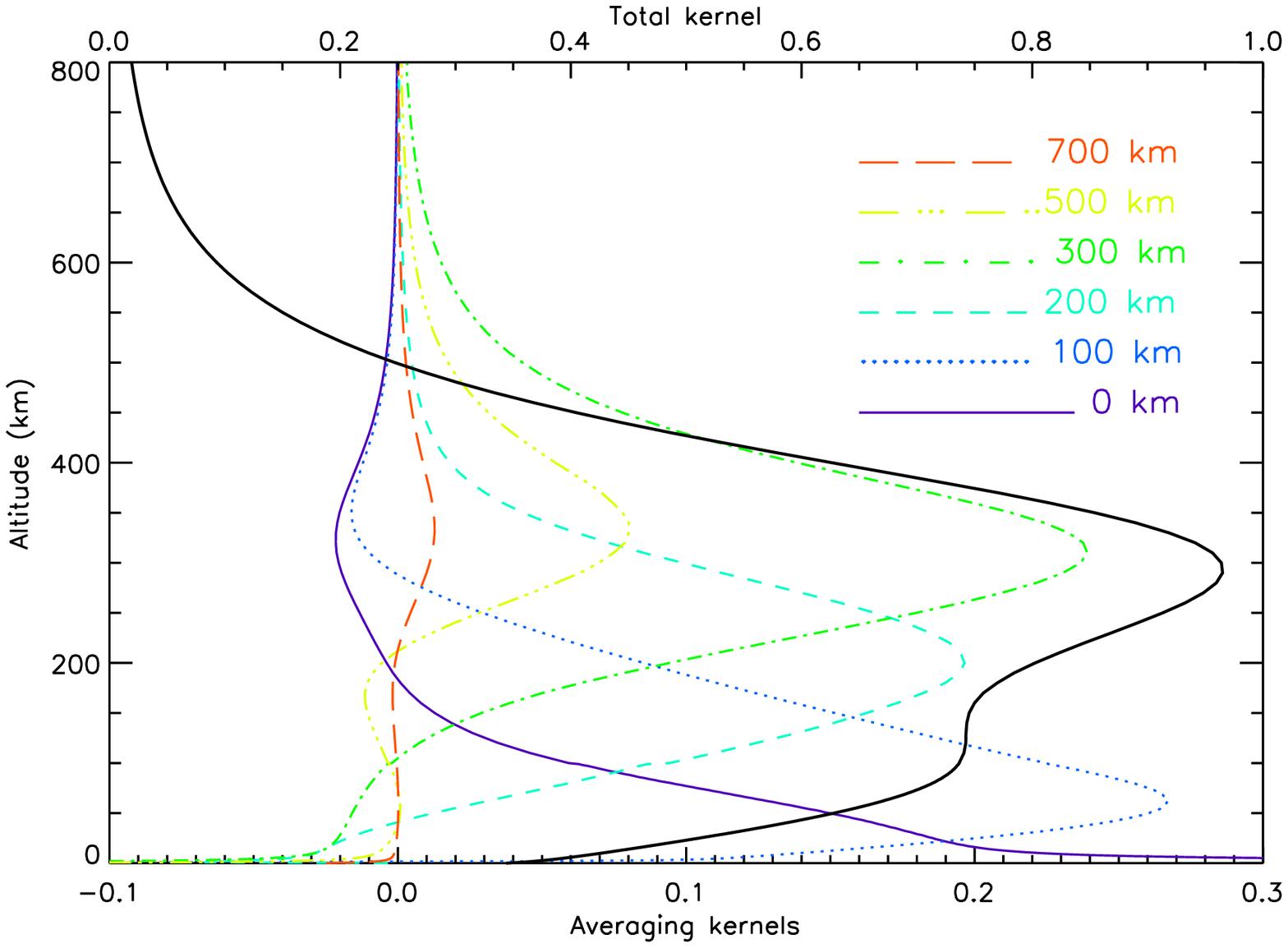,angle=0,width=15cm}

\vspace*{.5cm}
Fig. 4

\newpage
\epsfig{file=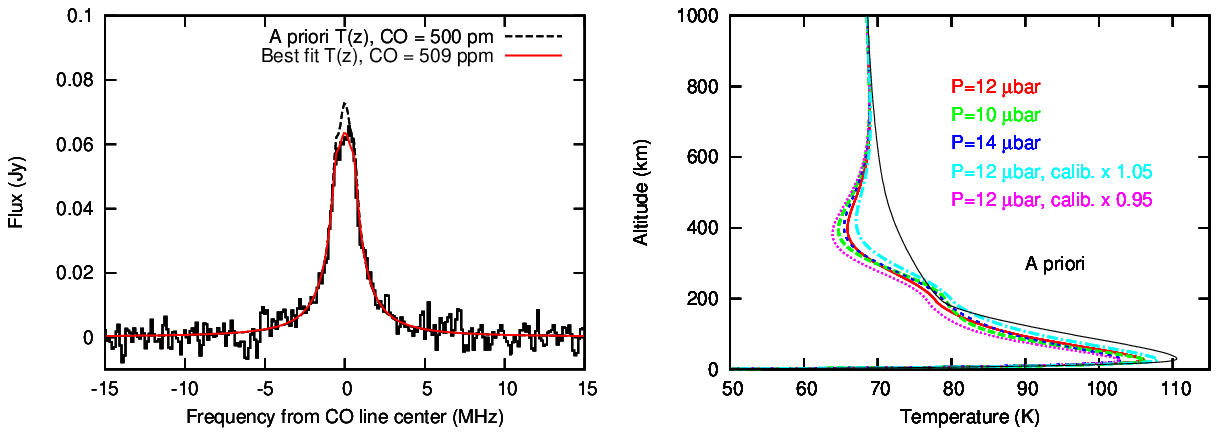,angle=-90,width=15cm}

\vspace*{.5cm}
Fig. 5

\newpage
\epsfig{file=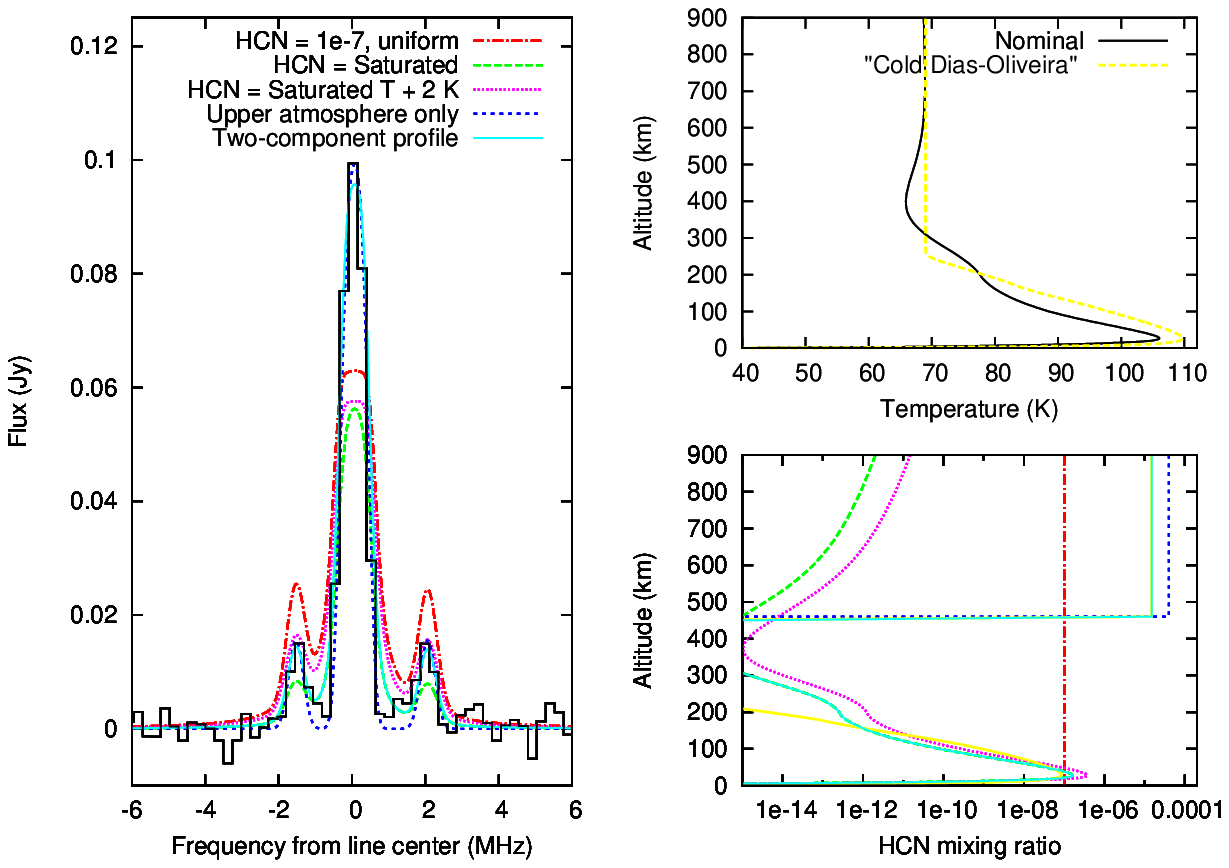,angle=0,width=15cm}

\vspace*{.5cm}

Fig. 6

\newpage
\epsfig{file=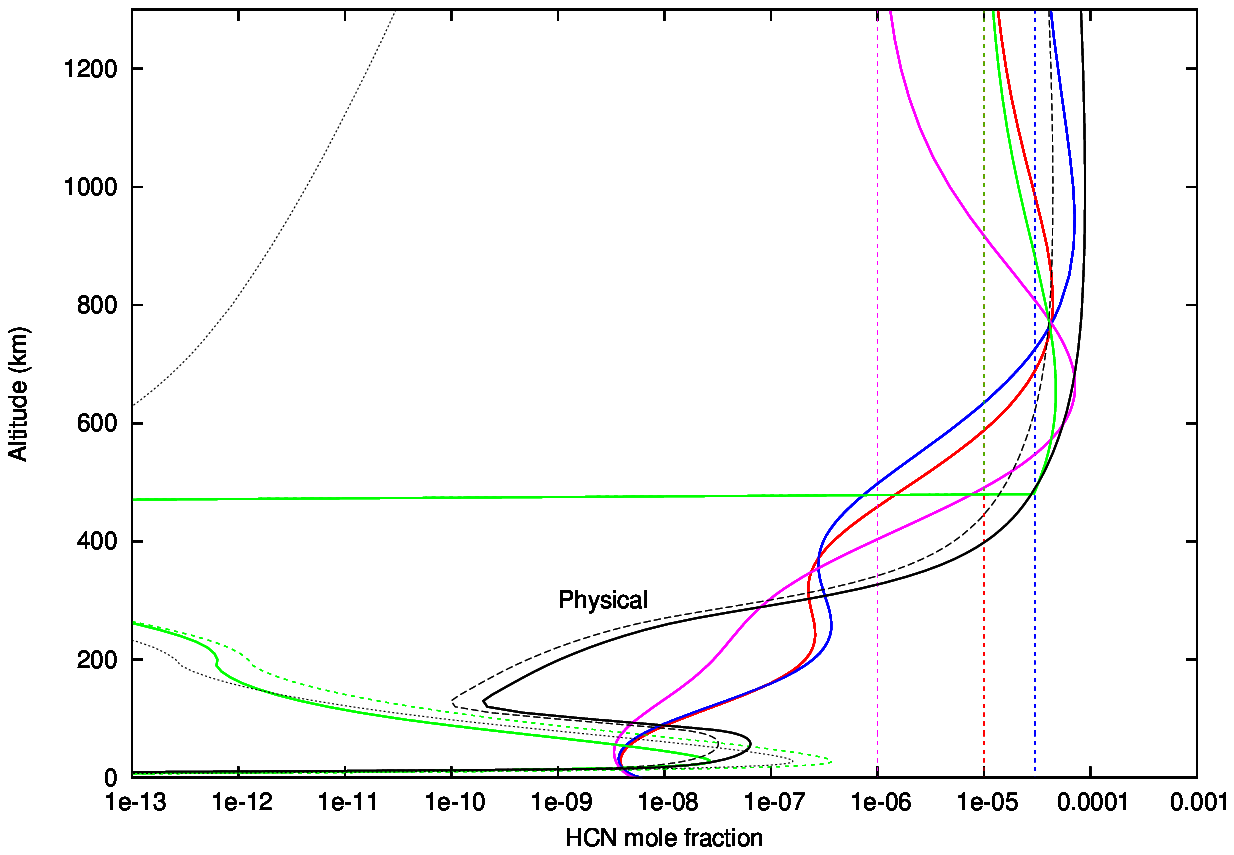,angle=-90,width=15cm}

\vspace*{.5cm}
Fig. 7

\newpage
\epsfig{file=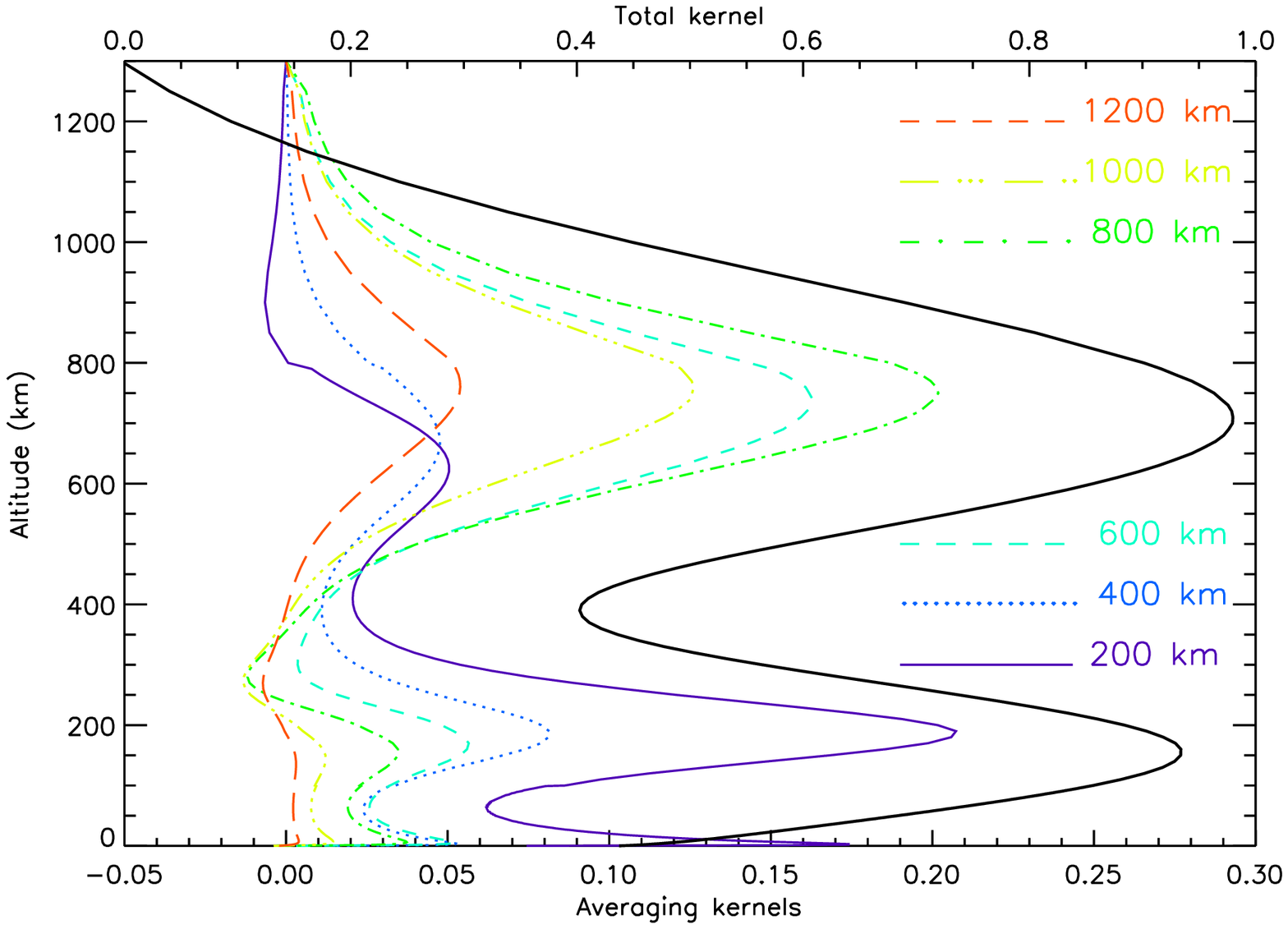,angle=0,width=15cm}

\vspace*{.5cm}
Fig. 8

\newpage
\epsfig{file=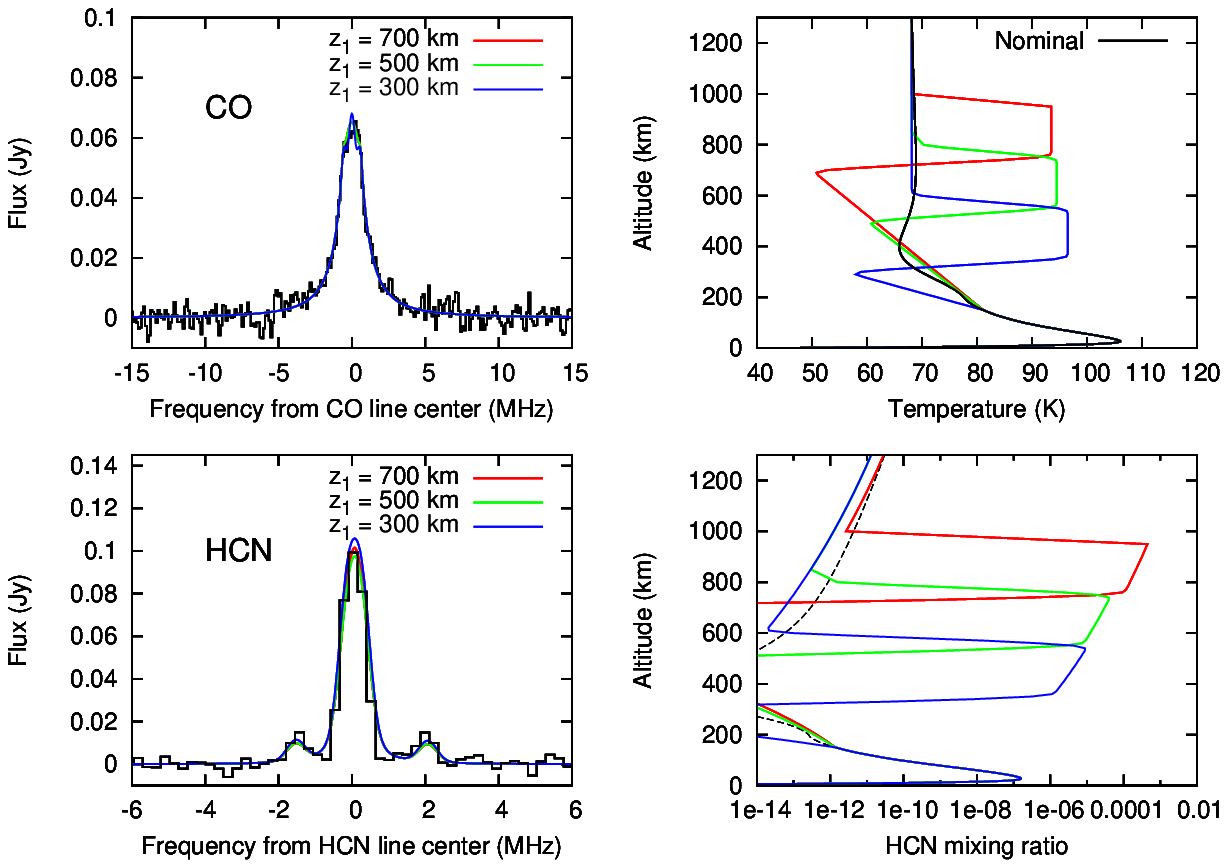,angle=-90,width=15cm}

\vspace*{.5cm}
Fig. 9

\newpage
\epsfig{file=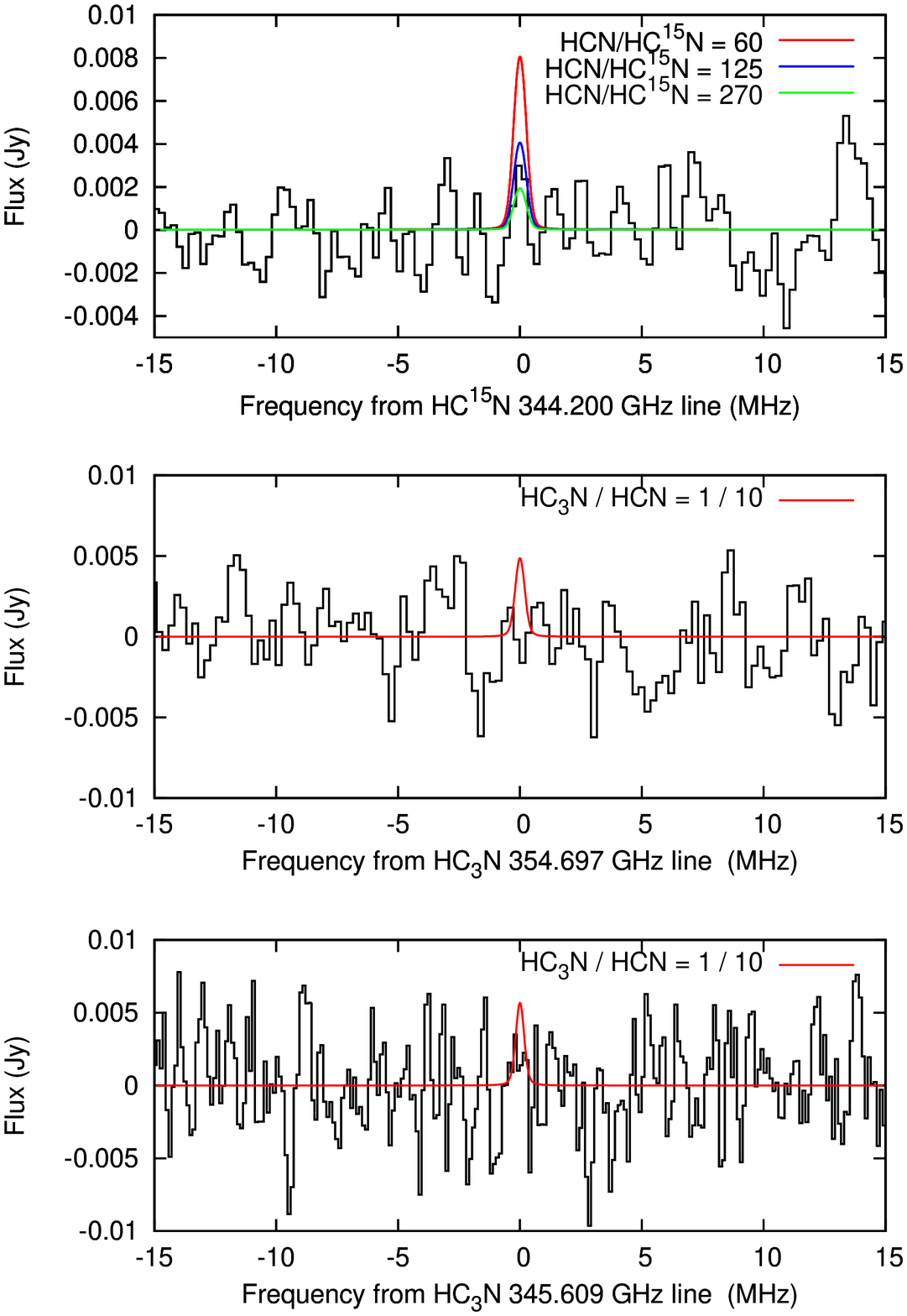,angle=0,width=15cm}

\vspace*{.5cm}
Fig. 10

\newpage
\epsfig{file=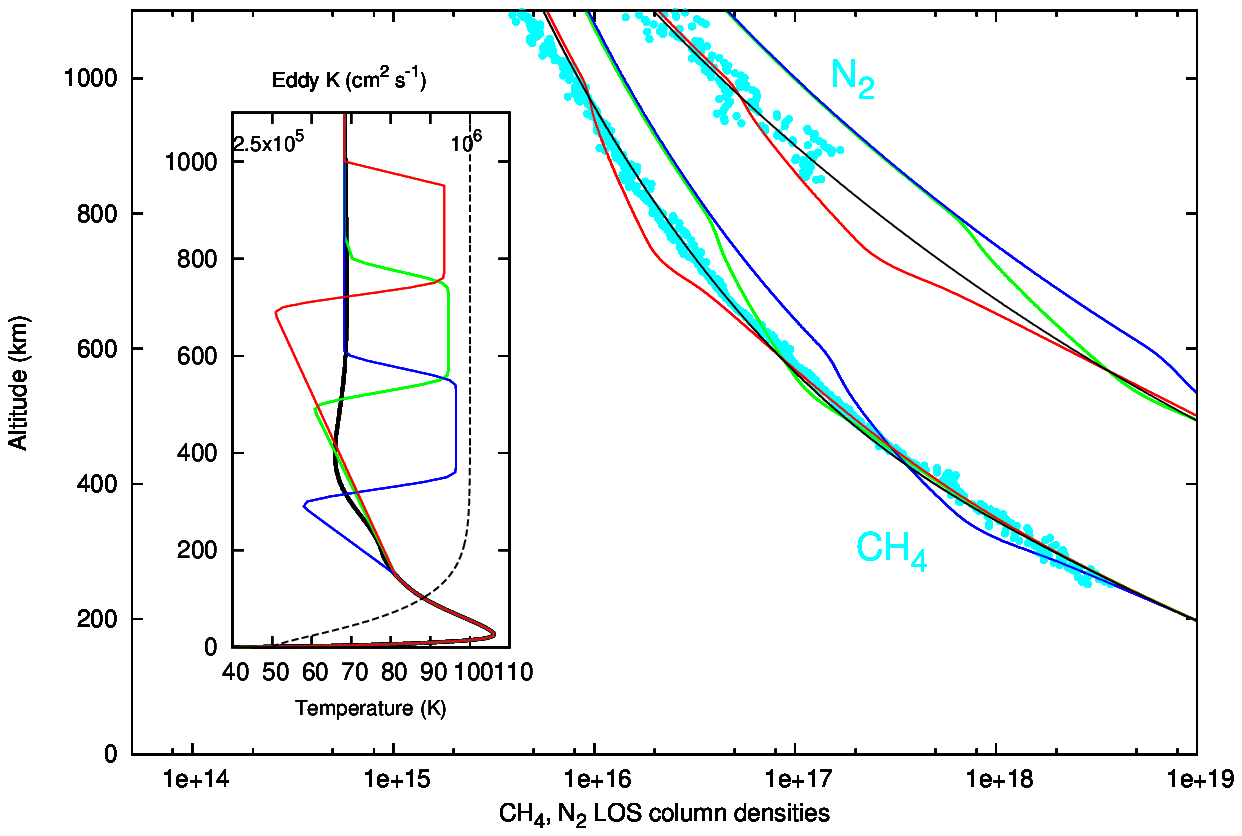,angle=-90,width=15cm}

\vspace*{.5cm}
Fig. 11

\newpage
\epsfig{file=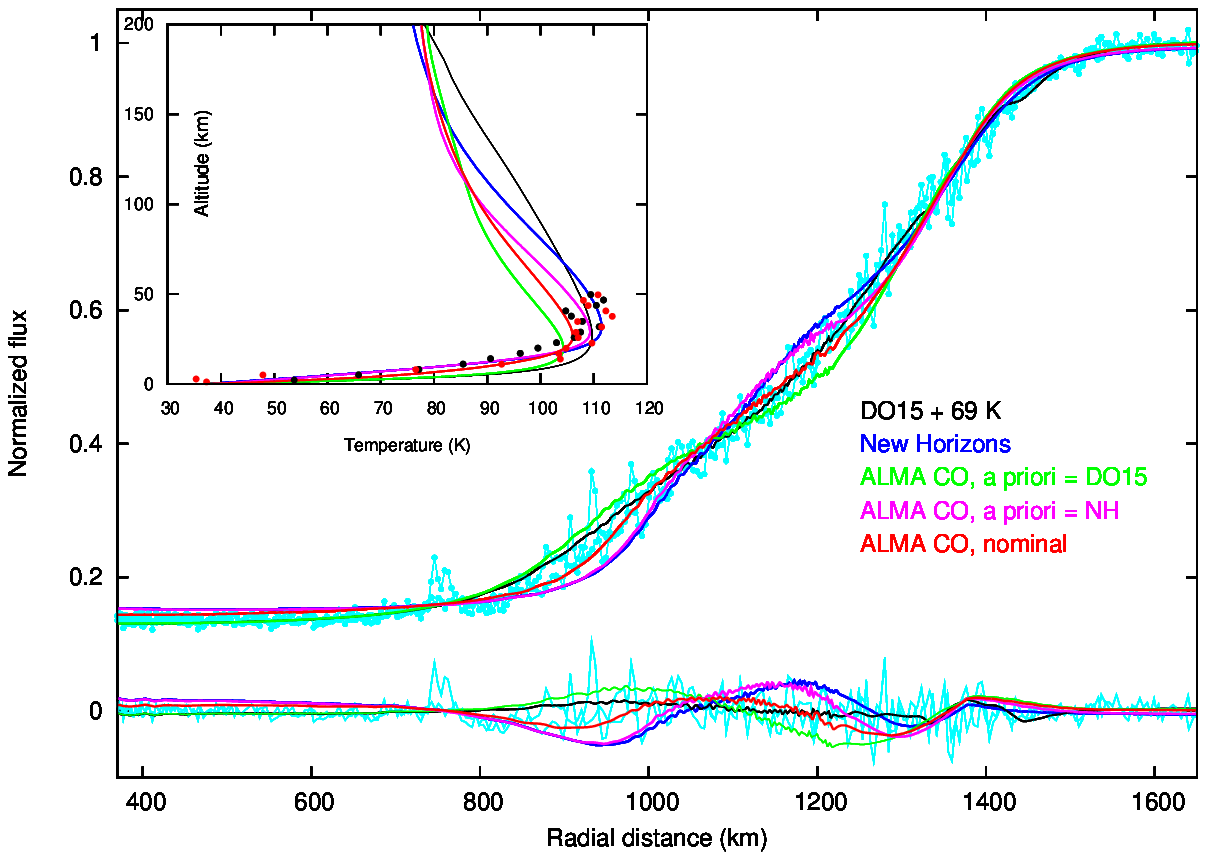,angle=-90,width=15cm}

\vspace*{.5cm}
Fig. 12

\newpage
\epsfig{file=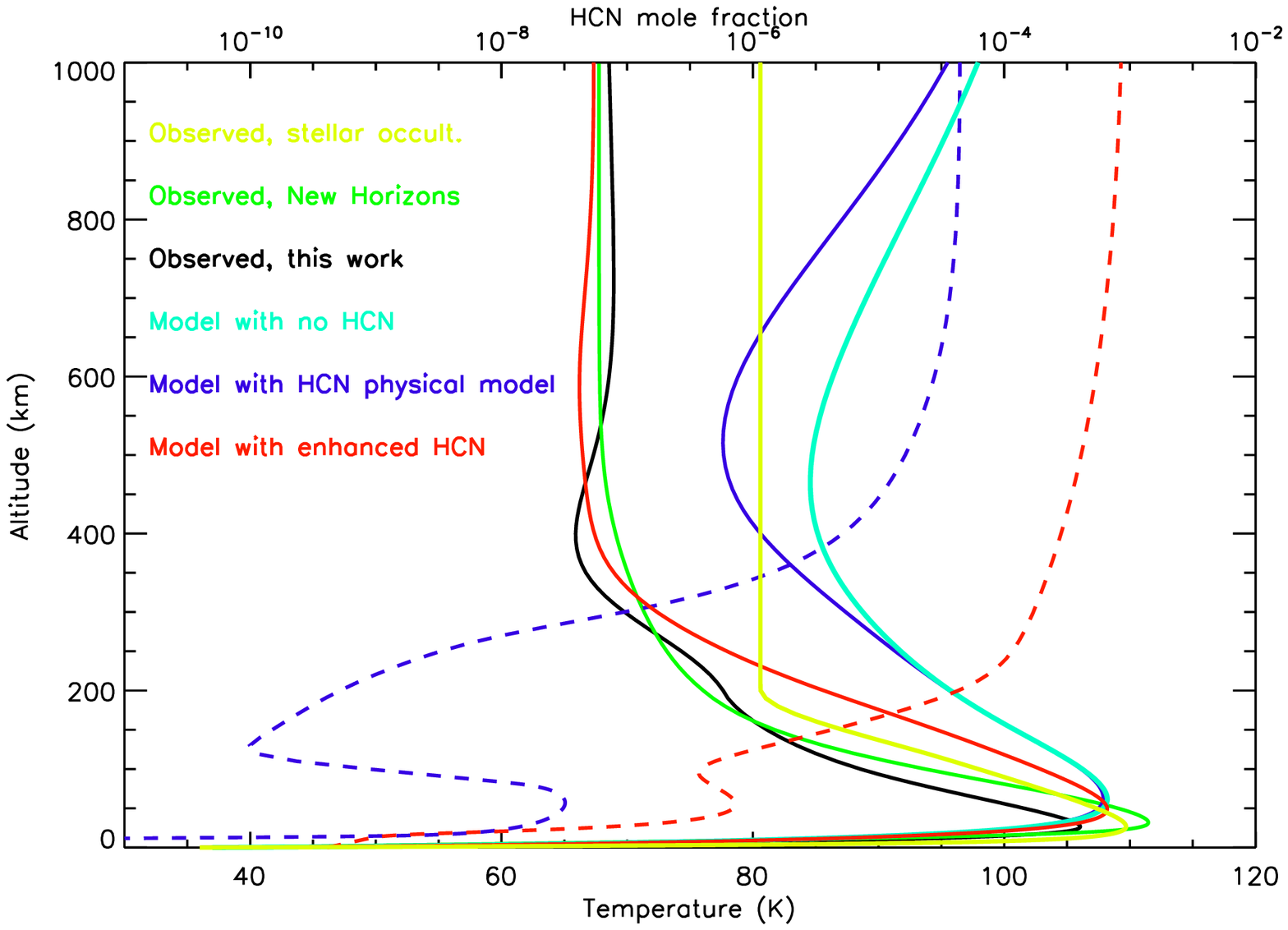,angle=0,width=15cm}

\vspace*{.5cm}
Fig. 13
 

\end{document}